\documentclass{article}
\pdfoutput=1

\usepackage{arxiv}

\usepackage{amsmath}
\DeclareMathOperator*{\argmax}{arg\,max}

\def\ci{\perp\!\!\!\perp}
\usepackage{algorithm}
\usepackage{algcompatible}
\usepackage{amsfonts}
\usepackage{amssymb}
\usepackage{amsbsy}
\usepackage[T1]{fontenc}
\usepackage{textcomp}
\usepackage{graphicx}
\usepackage{listings}
\usepackage[normalem]{ulem}
\usepackage{color}
\definecolor{gray}{rgb}{0.5,0.5,0.5}
\definecolor{brown}{rgb}{0.5,0.3,0}
\definecolor{ruby}{rgb}{0.6,0,0.3}
\definecolor{maroon}{rgb}{0.8,0,0.4}
\definecolor{rose}{rgb}{1.,0,0.4}
\definecolor{cream}{rgb}{1,0.95,0.8}
\definecolor{blueice}{rgb}{0.8,0.9,1}

\usepackage{graphicx}

\usepackage[hidelinks]{hyperref} 
\usepackage{fixltx2e}
\usepackage{soul}

\usepackage{tabularx}
\usepackage{longtable}
\usepackage{caption}
\usepackage{afterpage} 
\usepackage{colortbl}
\newcolumntype{Y}{>{\centering\arraybackslash}X}

\newcommand{\ignore}[1]{}
\newcommand{\revisit}[1]{}

\usepackage[utf8]{inputenc} 

\usepackage{booktabs}       
\usepackage{microtype}      

\title{Bayesian Surface Warping Approach For Rectifying Geological Boundaries Using Displacement Likelihood and Evidence from Geochemical Assays}
\shorttitle{Bayesian Surface Warping Approach}

\author{
  \textbf{Raymond~Leung}\thanks{Corresponding author. All authors are affiliated with the Rio Tinto Centre for Mine Automation and work in ACFR at The University of Sydney.}, \textbf{Alexander~Lowe},\\ \textbf{Anna~Chlingaryan}, \textbf{Arman~Melkumyan}, \textbf{John~Zigman}\vspace{2mm} \\
  Australian Centre for Field Robotics (ACFR)\\
  Faculty of Engineering\\
  The University of Sydney\\
  Sydney, NSW 2006 \\
  \texttt{raymond.leung@sydney.edu.au} \\
}

\renewcommand{\headeright}{Leung, Lowe et al.}

\begin{document}
\maketitle

\begin{abstract}
This paper presents a Bayesian framework for manipulating mesh surfaces with the aim of improving the positional integrity of the geological boundaries that they seek to represent. The assumption is that these surfaces, created initially using sparse data, capture the global trend and provide a reasonable approximation of the stratigraphic, mineralization and other types of boundaries for mining exploration, but they are locally inaccurate at scales typically required for grade estimation. The proposed methodology makes local spatial corrections automatically to maximize the agreement between the modelled surfaces and observed samples. Where possible, vertices on a mesh surface are moved to provide a clear delineation, for instance, between ore and waste material across the boundary based on spatial and compositional analysis; using assay measurements collected from densely spaced, geo-registered blast holes. The maximum a posteriori (MAP) solution ultimately considers the chemistry observation likelihood in a given domain. Furthermore, it is guided by an apriori spatial structure which embeds geological domain knowledge and determines the likelihood of a displacement estimate. The results demonstrate that increasing surface fidelity can significantly improve grade estimation performance based on large-scale model validation.
\end{abstract}

\keywords{Geochemistry-based Bayesian deformable surface (GC-BDS) model\and Bayesian Computation\and Mesh Geometry\and Surface Warping\and Spatial Correction\and  Displacement Likelihood\and Geological Boundaries\and Model Integrity.\newline\newline
\textbf{CCS Concepts}:\newline \hspace{8mm}$\bullet$ Computing methodologies\,$\rightarrow$\, Mesh geometry models;\newline$\bullet$ Mathematics of computing\,$\rightarrow$\, Bayesian computation;\newline $\bullet$ Applied computing\,$\rightarrow$\,Earth and atmospheric sciences.
}

\newpage\section{Introduction}\label{sec:bsu-intro}
Spatial structures and spatial algorithms are sometimes overshadowed by geostatistics and geochemical analysis in geology and stratigraphic modelling despite playing a no-less important role. In this paper, \textit{surface warping} is concerned with computational methods for mesh surface manipulation that automatically increase spatial fidelity \cite{guglielmino20133d}.\footnote{This should not be confused with alternative meanings in geomorphology for instance where \textit{surface warping} is attributed to cross-bending stresses resulting from torsional forces in the study of the mechanics of geologic structures \cite{mead1920notes}.} Specifically, it focuses on reshaping and correcting inaccuracies in an existing surface to maximize its agreement with observed data; in particular, geochemical assays sampled from drilled holes in an ore deposit.

For decades, displacement field estimation (e.g. \cite{krishnamurthy1995optical}\cite{chang1997simultaneous}\cite{mark1997post}\cite{farneback2001very}\cite{secker2003lifting}) and surface editing techniques (e.g. \cite{nealen2006physically}\cite{sorkine2004laplacian}) have flourished in the computer vision and computer graphics community. The idea of directly manipulating mesh surface vertices can be traced back to Allan, Wyvill and Witten \cite{allan1989methodology} amongst others in a contemporary context, where the concepts of region of influence, movement constraints (in terms of decay function, bound and anchored vertices) are discussed.

In specific disciplines such as video coding, dense motion field (optical flow) estimation and segmentation have, for instance, been attempted using deformable mesh motion models and MAP estimators. The a priori distribution of the estimates can be modelled by a coupled Markov random field to account for both spatial smoothness and temporal continuity along the estimated motion trajectories \cite{stiller1994object}. Despite these similarities, the \textit{surface and boundary alignment} problem considered in this paper differs in some significant ways. Although displacement estimation remains a central theme, the observations rely on \textit{geochemistry} rather than photogrammetry (or interferometry in the case of strain tensor estimation from geodetic and satellite deformation measurements \cite{guglielmino20133d}) and these observations are sparse, spatially irregular and noisy in comparison.

For the surface and boundary alignment problem, mesh processing techniques \cite{botsch2010polygon} and domain knowledge \cite{dewaele-18} can play a crucial role. To the authors' knowledge, \textit{displacement field estimation} has not been utilized previously in 3D surface-based modelling of geological structures \cite{caumon2009surface} or exploited in practice to improve the efficacy of models in mining. This paper aims to bridge the gap that exists between a model and the latest data acquired from a mine. The Bayesian approach basically connects a model (or belief) with evidence from geochemical observations (the reality). Displacement estimation helps identify discrepancies, then surface warping corrects spatial inaccuracies in the surface-based representation of the relevant geological structures.

In the literature, research has progressed largely in two separate streams: (1) advances in computational techniques that manipulate surfaces using dense uniformly sampled data, where warping is used to process visual information or register anatomical changes via MRI \cite{thompson2000warping} for instance; (2) generating 3D subsurface models using extensive field data and different modalities, a variety of techniques (including the use of contours and differential  geometry) are summarized in \cite{leung-19subsurface} (Sect.\,1.1) and \cite{caumon2009surface}.

In the first research stream, examples include (a) energy constrained mesh deformation using subspace gradient domain techniques \cite{huang2006subspace} where the targeted application is computer animation; and (b) post-rendering 3D warping which compensates for occlusion-related artifacts due to viewpoint variation by compositing interpolants from multiple reference frames \cite{mark1997post}. In computer graphics, methods for constructing extrinsically smooth and globally optimal directional fields have been considered by Jakob et al.\,\cite{jakob2015instant}, Huang and Ju \cite{huang2016extrinsically} and Kn{\"o}ppel et al.\,\cite{knoppel2013globally} where it is treated as a discrete optimization or sparse eigenvalue problem.

An example from the second stream is the work of Olierook et al.\,\cite{olierook2021bayesian} which uses a Bayesian approach to fuse lithostratigraphic field observations with aeromagnetic and gravity measurements to build 3D geological models without structural data. Such model is fit for purpose for mineral exploration; however the grade control requirements encountered in a mine production setting are generally more stringent and demands far greater precision and vertical resolution. In \cite{calcagno2008geological}, Calcagno et al.\, use cokriging \cite{ver1998constructing} to create a continuous 3D potential field to describe the geometry of the geology. The model input consists of the location of geological interfaces and orientation data. Subsequently, structural information such as the location and orientation of boundaries are extracted from isosurfaces and the gradient of the interpolated potential function.

The state of the art \cite{maccormack-19} generally considers modelling as an open-loop process where the input data is complete and a static model is to be produced. In practice, data is usually harvested bench-by-bench in a piecemeal manner in open-pit mining. Thus, there is a strong incentive in utilizing newly acquired data to improve both surface definitions and the existing model, to further understand the subterranean deposit below the current bench for mine planning and operational guidance.

The missing link is a synergy between geochemical data, surface representation and the model --- and whether grade prediction performance can benefit from an incremental update strategy. These novel issues have not received much attention in earth and spatial science. In terms of where this paradigm fits, based on Nealen's survey article \cite{nealen2006physically}, a new category called geochemistry-based Bayesian deformable surface (GC-BDS) model is proposed to characterize the warping approach presented in this paper.

\subsection{Motivation}\label{sec:motivation}
To understand the importance of surface and boundary alignment, we first consider the implications of working with an inaccurate (or misleading) surface and its flow-on effects on subsequent modelling processes before formulating a Bayesian framework for surface warping with a view of improving surface integrity. As motivation, we illustrate how starting with a bad surface --- one that is not representative of the true geological boundary --- can impact on the block structure and inferencing ability of a grade estimation model.

\begin{figure*}[!htb]
\centering
\includegraphics[width=140mm]{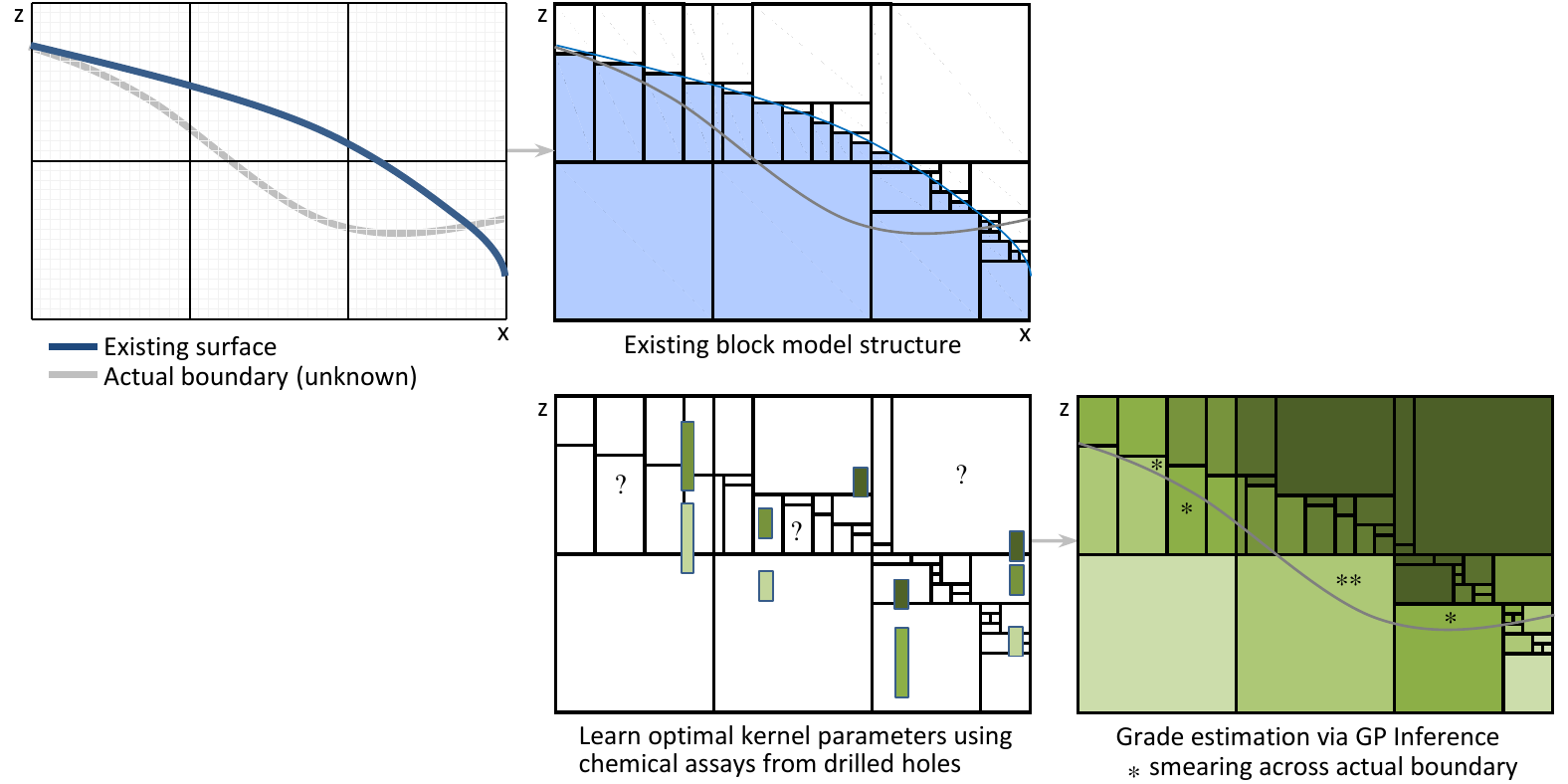}
\caption{Computational pipeline for the orebody grade estimation task. Block model spatial restructuring and inferencing both suffer from the negative effects of a surface that misrepresents the true geological boundary. The question marks serve as a reminder that the grade value of all the blocks need to be estimated.}
\label{fig:warping-motivation1}
\end{figure*}

Figure~\ref{fig:warping-motivation1} illustrates the computational pipeline for a typical grade estimation problem where the main objectives are (i) obtain a compact block-based representation of the orebody with good boundary localization property; (ii) estimate the grade value for chemical components of interest, predicting these values especially at locations where no measurements are available using geostatistical or machine learning techniques. For the first objective, it requires changing the spatial structure of the model, partitioning the blocks as necessary (down to some predetermined, acceptable minimum block size) to closely follow the location and approximate the curvature of the geological boundary which the surface seeks to represent. In Fig.~\ref{fig:warping-motivation1}(top middle), we have a situation where the existing surface misrepresents the location of the actual boundary [which is not directly observable].\footnote{For illustrative purpose, Fig.~\ref{fig:warping-motivation1} shows a situation where the modelled boundary is grossly misplaced. In practice, the discrepancies that exist between the actual and initial modelled boundary are often explained by local differences due to inadequate sampling.}

Consequently,  the spatial restructuring algorithm \cite{leung2020mos} produces a block structure that is not faithful to the underlying boundary through no fault of its own. For the second objective, using a Gaussian Process (GP) learning and inferencing approach [described in Appendix~\ref{sect:inferencing-for-grade-estimation}], spatial variations (covariance functions) are learnt using samples drawn from an incorrect geological domain structure. A major consequence, depicted in Fig.~\ref{fig:warping-motivation1}(bottom right), is that the inferenced values for the blocks are biased near the actual boundary due to under-estimation or over-estimation of the chemistry values, and the blocks not properly decomposed to follow the shape of the boundary. These two factors lead to smearing in the predicted chemistry across the actual boundary which has practical implications for ore extraction and material blending in a production setting.

\begin{figure*}[!htb]
\centering
\includegraphics[width=140mm]{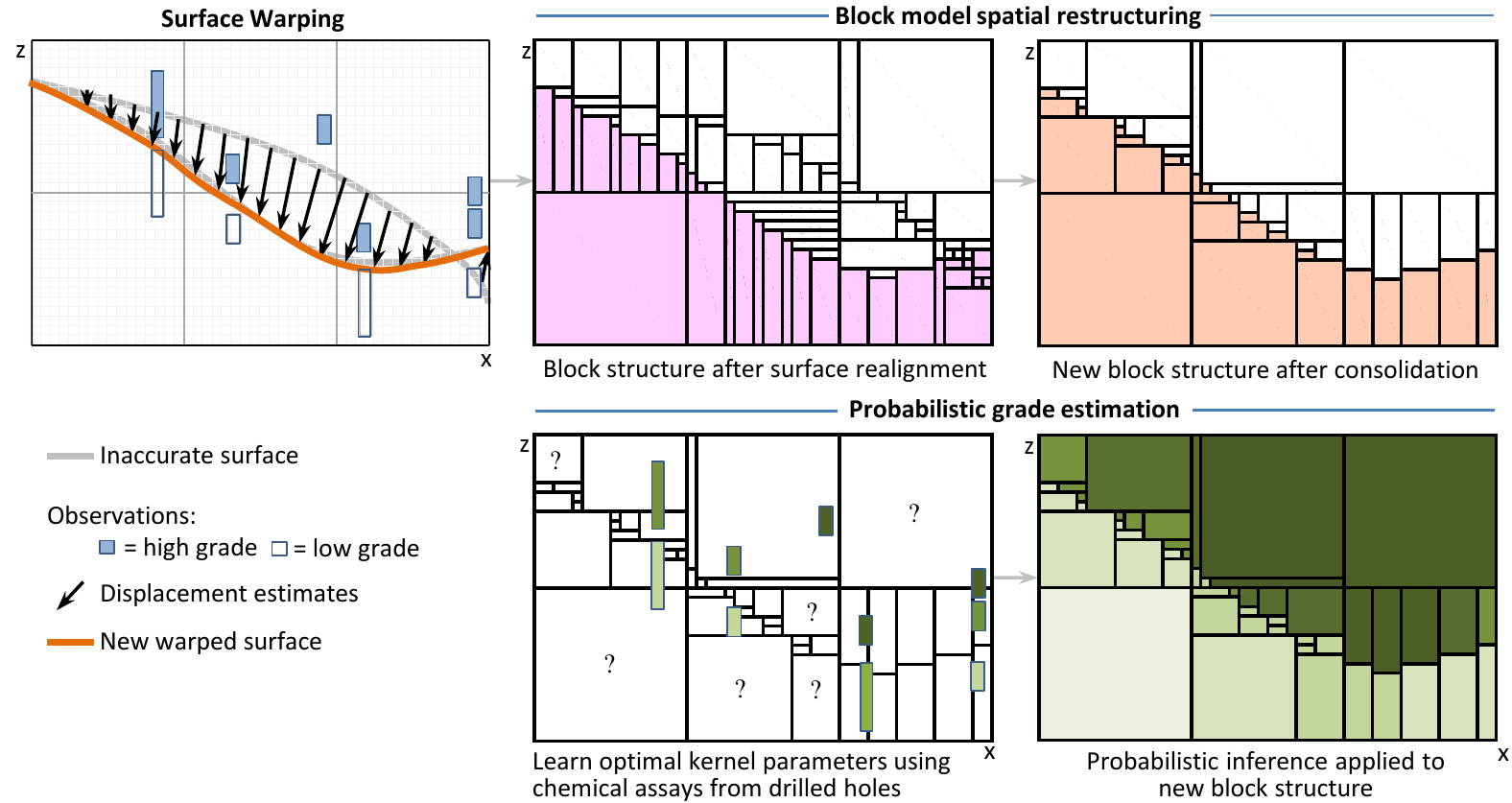}
\caption{Rectification of the boundary through surface warping improves the spatial structure and inferencing ability of the grade block model.}
\label{fig:warping-motivation2}
\end{figure*}

The main proposition of this paper is a Bayesian approach that corrects inaccuracies in a mesh surface through spatial warping which thus increases the positional integrity of the underlying boundary that it seeks to represent. Conceptually, the problem involves estimating the displacements to bring an existing surface into alignment with the true boundary based on the observed sample chemistry. The aim is to capture more precisely the shape and location of local features; this is depicted in Fig.~\ref{fig:warping-motivation2}(top left). The existing block structure is modified adaptively to align itself with the new warped surface. In Fig.~\ref{fig:warping-motivation2}(top center), the new block structure evidently follows the curvature of the new surface, however some of the preexisting (subdivided) blocks from the earlier misplaced boundary still remain. A block merging algorithm proposed in \cite{leung2020mos} is used to coalesce fragmented blocks to produce the consolidated block structure shown in Fig.~\ref{fig:warping-motivation2}(top right). The GP inferencing procedure is applied once again to the new block structure. The result shown in Fig.~\ref{fig:warping-motivation2}(bottom right) is free of blending or smearing artefacts. When the surface is accurate, the resultant model exhibits a clear contrast in the predicted grade values on either side of the actual boundary. Improving the reliability of these estimates ultimately enables better decision making, planning and ore extraction.

To summarise, an ill-placed boundary can impact the block structure and grade estimation in significant ways. The boundary localization properties deteriorate when a block model is partitioned by an inaccurate surface. Incorrect geological domain classification resulting from a misleading surface can cause smearing to occur during inferencing where compositions are over-estimated or under-estimated near the actual boundary.

\subsection{Contributions}\label{sect:contributions}
In Section~\ref{sect:surface-warping}, the surface warping problem is formulated in a Bayesian framework and the MAP (maximum a posteriori) solution for node displacements estimation is presented; this serves to maximize the agreement between surfaces and geochemical observations from blast hole samples. In Section~\ref{sect:boundary-warping-fusion-workflow-validation}, local improvements are visually highlighted. An objective measure called \textit{r\textsubscript{2} cdf error score} is proposed and used for model validation to demonstrate an improvement in grade estimation performance resulting from spatial warping.

\section{Surface warping}\label{sect:surface-warping}
\begin{figure*}[!htb]
\centering
\includegraphics[width=85mm]{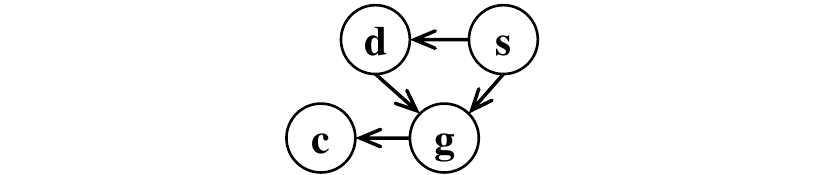}
\caption{Graphical model for surface warping. The graph expresses the conditional dependence structure between the random variables: \textbf{c}=observed chemistry, \textbf{g}=geozone, \textbf{d}=displacement and \textbf{s}=spatial properties. The arrows represent conditional dependency of a target node from a source node.}
\label{fig:graphical-model}
\end{figure*}

Surface warping may be framed as a Bayesian parametric inference problem where the objective is to estimate the required displacements (or spatial corrections) $\boldsymbol{\theta}\equiv \mathbf{d}\in\mathbb{R}^3$ to maximize the positional integrity of geological boundaries given a set of observations. In general terms, the observations consist of the location $\mathbf{x}\in\mathbb{R}^3$ and spatial extent $\boldsymbol
{\delta}=[0,0,h]^T\subseteq\mathbb{R}^3$ of the measurements, as well as the composition $\mathbf{c}\in\mathbf{R}^K$ of the sample determined by chemical assays. The prior information available is a reference geological structure $\mathcal{G}$ that determines the geozone classification ($g\in\mathbb{Z}$) of a sample which indicates the geological domain where it belongs. This prior embeds knowledge about the stratigraphic structure of the modelled region which is considered a faithful (unbiased) representation of the ground truth at large scales, but inaccurate at smaller scales ($\sim 5-20$m) due to sparse sampling and local variation. Accordingly, applying Bayes rule, the problem may be formulated as
\begin{align}
\argmax_{\mathbf{d}} P(\mathbf{d}\!\mid\! \mathbf{c},\mathbf{s})\label{eq:surface-warping-problem}
\end{align}
where spatial information is contained in $\mathbf{s}=[\mathbf{x},\boldsymbol{\delta}]^T\in\mathbb{R}^6$ and
\begin{align}
P(\mathbf{d}\!\mid\! \mathbf{c},\mathbf{s})&\propto P(\mathbf{c}\!\mid\!\mathbf{d},\mathbf{s})\, P(\mathbf{d}\!\mid\!\mathbf{s}) \label{eq:surface-warping-bayes-formula1}\\
&= \left(\sum_{g} P(\mathbf{c},g\!\mid\! \mathbf{d},\mathbf{s})\right) P(\mathbf{d}\!\mid\!\mathbf{s}) \label{eq:surface-warping-bayes-formula2}\\
&= \left(\sum_{g} P(\mathbf{c}\!\mid\! g,\mathbf{d},\mathbf{s}) P(g\!\mid\! \mathbf{d},\mathbf{s})\right) P(\mathbf{d}\!\mid\!\mathbf{s}) \label{eq:surface-warping-bayes-formula3}\\
&= \sum_{g} P(\mathbf{c}\!\mid\! g,\mathbf{d},\mathbf{s}) P(g\!\mid\! \mathbf{d},\mathbf{s}) P(\mathbf{d}\!\mid\!\mathbf{s}) \label{eq:surface-warping-bayes-formula4}\\
&\approx \sum_{g} P(\mathbf{c}\!\mid\! g) P(g, \mathbf{d}\!\mid\! \mathbf{s})\label{eq:surface-warping-bayes-formula5}
\end{align}
Marginalization, conditional probabilities and conditional independence,\footnote{In our experience, using $p(\mathbf{c}\!\mid\! g,\mathbf{d},\mathbf{s})$ does not further increase performance since spatial attributes are already modelled by $P(g,\mathbf{d}\!\mid\!\mathbf{s})$.} ($\mathbf{c}\ci \mathbf{d},\mathbf{s})\mid g$, are used in (\ref{eq:surface-warping-bayes-formula2}), (\ref{eq:surface-warping-bayes-formula3}) and (\ref{eq:surface-warping-bayes-formula5}), respectively. The final expression in (\ref{eq:surface-warping-bayes-formula5}) offers a clear interpretation for the posterior $P(\mathbf{d}\!\mid\! \mathbf{c},\mathbf{s})$. Specifically, $P(\mathbf{c}\!\mid\! g)$ denotes the likelihood of observing the chemical composition $\mathbf{c}$ in geozone $g$, whereas $P(g, \mathbf{d}\!\mid\! \mathbf{s})$ represents a spatial prior that considers the displacement and geozone likelihood given the sample location. This Bayesian network is described by the graphical model shown in Fig.~\ref{fig:graphical-model}. Even for this simple structure, there is tremendous scope for ingenuity. The following describes one possible implementation and reflects on some practical issues.

For readers new to geostatistics, a common pitfall is an attempt to model $P(\mathbf{c}\!\mid\! g)$ directly using raw chemical assay measurements and an affinity measure such as the Mahalanobis distance, $d^2(\mathbf{c}^\text{sample},\mathbf{c}^\text{class})$. As Filzmoser et al.\cite{filzmoser-hron-08} have pointed out, statistical analysis performed on compositional data\footnote{A key characteristic of compositional data is that they lie in the Aitchison Simplex \cite{filzmoser-hron-08}, which means the data is not Gaussian (or even symmetrically) distributed. An increase in a key component may cause another to decrease and the components sum to a constant.} without resorting to log-ratio transformation may lead to invalid or dubious interpretations \cite{garrett-17}. Even when the Mahalanobis distance metric can be legitimately applied following isometric log-ratio transformation \cite{greenacre-19}, it may still be inappropriate, as errors of the same magnitude may have different levels of significance depending on the grading thresholds. In grade estimation work, samples are routinely categorized based on composition. Thus, instead of $P(\mathbf{c}\!\mid\! g)$, a likelihood probability mass function $L(y(\mathbf{c})\!\mid\! g)$ is used by Lowe in \cite{rtgi-rtcma2018} where $y(\mathbf{c})$ represents a categorical label. These labels generally correspond to mineralogical groupings or `destination tags' since the excavated materials will be sorted eventually based on chemical and material properties.\footnote{For downstream ore processing, it is useful to know whether the material is hard or friable, lumpy or fine, viscous or powdery. Typically, 6 to 12 categorical labels, $y(\mathbf{c})$, are used. For our purpose, we focus more on sample chemistry. The criteria for HG (high grade) and LGA (low grade aluminous) iron ore, for instance, might be set at Fe $\ge$ 60\% and ($50\%\le \text{Fe} < 60\%$, Al\textsubscript{2}O\textsubscript{3} $\ge 3\%$), respectively. These parameters vary depending on the deposit and geozone.}

In practice, there is a finite number of mineralogical groupings and geozones. Hence, $L(y(\mathbf{c})\!\mid\! g)$ is computed from a table with dimensions ($N_\text{class},N_\text{geozone}$) constructed using frequency counts applied to assay samples collected from exploration drillings. Here, $\mathbf{c}$ is observed, $y(\mathbf{c}): \mathbb{R}^K\rightarrow \mathbb{Z}$ is a deterministic mapping and $g$ is known. 

For the prior $P(g, \mathbf{d}\!\mid\! \mathbf{s}(\mathbf{x},\boldsymbol{\delta}))$, a proxy function $R(\mathbf{x}+\mathbf{d}, g, \boldsymbol{\delta})$ is used to compute the geozone and displacement likelihood in \cite{rtgi-rtcma2018}. This utilizes the a priori geological structure $\mathcal{G}$ to assess the feasibility of displacement $\mathbf{d}$. In particular, $R(\mathbf{x}+\mathbf{d}, g, \boldsymbol{\delta})$ determines the amount of overlap, $r\in[0,1]$, between geozone $g$ and an interval observation $\boldsymbol{\delta}$ of length $h$ at the proposed location $(\mathbf{x}+\mathbf{d})$. Alternatively, spatial correlation may be modelled using autocorrelation functions and random field simulation as demonstrated in \cite{gong2020stratigraphic} in regions where stratigraphic association can be reasonably inferred from dense drill hole samples. It is clear that a number of strategies can be used to find the optimal displacement in a hierarchical search space. For instance, a conditional random field may be used to impose connectivity and regularization constraints. For simplicity and speed, a set of candidate displacement points $\{\mathbf{d}_i\}$ may be chosen from a regular 3-D lattice in the vicinity of $\mathbf{x}$, viz., $\mathcal{L}_{\mathbf{x}}$. Assuming $\lvert \mathcal{L}_{\mathbf{x}}\rvert = N_\text{displacement}$ for all $\mathbf{x}$, the posterior would result in a table of size $(N_\text{sample},N_\text{displacement})$. The maximum a posteriori (MAP) estimate is given by (\ref{eq:surface-warping-max-posterior-estimate}), the solution with minimum $\lVert\mathbf{d}\rVert$ is chosen in the event of a tie.
\begin{align}
\mathbf{d}_\text{MAP}(\mathbf{x})&=\argmax_{\mathbf{d}} L(\mathbf{d}\!\mid\! y(\mathbf{c}),\mathbf{s}(\mathbf{x},\boldsymbol{\delta}))\notag\\
&= \argmax_{\mathbf{d}} \sum_g L(y(\mathbf{c})\!\mid\! g)R(\mathbf{x}+\mathbf{d}, g, \boldsymbol{\delta})\label{eq:surface-warping-max-posterior-estimate}
\end{align}

Diffusion flow techniques (based on using discrete Laplace-Beltrami \cite{botsch2010polygon}) may be applied to manifold surfaces to obtain a coherent displacement field where $\mathbf{d}_\text{MAP}(\mathbf{x})$ varies smoothly. However, dithering often presents as a simpler alternative. The solution is obtained as an aggregate average over a small neighbourhood, $\{\mathbf{x}+\boldmath{\epsilon}_i\}\in\mathcal{N}_\mathbf{x}$,\footnote{Another option is to treat $\mathcal{N}_\mathbf{x}$ as the barycentric cell or mixed Voronoi cell \cite{botsch2010polygon} in the 1-ring neighbourhood of $\mathbf{x}$.} with higher weights $w(\mathbf{x}_i)$ given to nearby estimates and displacements perpendicular to the surface normal $\mathbf{n}_\mathbf{x}$.
\begin{align}
\mathbf{d}'_\text{MAP}(\mathbf{x})=\sum_{\mathbf{x}_i\in\mathcal{N}_\mathbf{x}} w(\mathbf{x}_i)\cdot \mathbf{d}_\text{MAP}(\mathbf{x}_i)\label{eq:surface-warping-map-aggregate}
\end{align}
For instance, setting $w(\mathbf{x}_i)$ to $\text{proximity}(\mathbf{x},\mathbf{x}_i)\times\left|\left<\mathbf{n}_\mathbf{x},\mathbf{d}_\text{MAP}(\mathbf{x}_i)\right>\right|$ provides local smoothing and discourages movement parallel\footnote{Tangential movements do not effectively compensate for displacement errors which are perpendicular to the surface.} to the surface which is not productive.\footnote{In our implementation, IDW (inverse distance weights) are used for proximity($\mathbf{x},\mathbf{x}_i$). Another option is to use normalized exponential (softmax) function $\frac{e^{-\beta z_i}}{\sum_{i\in\mathcal{N}}e^{-\beta z_i}}$ where $z_i=\lVert\mathbf{x}-\mathbf{x}_i\rVert$. For the direction penalty term, $1 - (1 - {\cos}^2\theta)^2$ is used in place of $\left|\left<\mathbf{n}_\mathbf{x},\mathbf{d}_\text{MAP}(\mathbf{x}_i)\right>\right|$, where $\cos\theta = \left<\mathbf{n}_\mathbf{x},\mathbf{d}_\text{MAP}(\mathbf{x}_i)\right>$.} In instances where stratigraphic forward modelling (SFM) hints are available, $\mathbf{n}_\mathbf{x}$ may be guided instead by directional projections based on the deposition and evolution of sedimentary facies within a stratigraphic framework \cite{huang2015recent}. Given a set of mesh surface vertices $\mathbf{x}_q\in\mathcal{S}$, their corrected positions after surface warping are given by
\vspace{-1mm}\begin{align}
\mathbf{x}'_q=\mathbf{x}_q-\mathbf{d}'_\text{MAP}(\mathbf{x}_q).\label{eq:surface-warping-vertices-correction}
\end{align}
Equation (\ref{eq:surface-warping-vertices-correction}) simply applies the displacement-error corrections to surface vertices.

Computing this expression usually requires spatial interpolation as $\mathbf{d}_\text{MAP}(\mathbf{x})$ is initially evaluated at sparse locations where assay information (geochemical evidence) is available. In areas where spatial resolution is low, mesh surface triangles may be subdivided to increase point density. The classification function, $y(\mathbf{c}): \mathbb{R}^K\rightarrow \mathbb{Z}$, and more generally $p(\mathbf{c}\!\mid\! g)$, may be learned \cite{leung2021warpml} using supervised or unsupervised techniques when the rules for destination tags (mineralogical grouping) are inadequate or unavailable.

\subsection{Algorithm}\label{sect:spatial-warping-algorithm}
The Bayesian surface warping algorithm may be summarised in a series of steps. The description here is immediately followed with explanation and illustration of the key steps in Sec.~\ref{sect:illustration}.
\begin{enumerate}
\item[]\hspace{-9mm}Given a chemical assay to material type categorical mapping $y(\mathbf{c}): \mathbb{R}^K\rightarrow \mathbb{Z}$,
\item Compute $L(y(\mathbf{c})\!\mid\! g)\in\mathbb{R}^{N_\text{class}\times N_\text{geozone}}$ using training samples $\{(\mathbf{c}_j, g_j)\}_j$ from exploration holes\label{algo:step1}
\item[]\hspace{-9mm}For each blast hole sample $i=1,\ldots, N_\text{sample}$:
\item Assign categorical label $y(\mathbf{c}_i)\in\{1,\ldots,N_\text{class}\}$ to each sample\label{algo:step2}
\item Compute $L(y(\mathbf{c}_i)\!\mid\! g)$ across all $N_\text{geozone}$ geozones\label{algo:step3}
\item Compute geozone-displacement likelihood, $R(\mathbf{x}_i+\mathbf{d}_i, g,\boldsymbol{\delta}_i)$\label{algo:step4}
\item[]\hspace{-9mm}For each surface vertex $\mathbf{x}_q$ and candidate displacement vector $\mathbf{d}_k$ from $\mathcal{L}_{\mathbf{x}_q}, k\in\{1,\ldots,N_\text{displacement}\}$:
\item Interpolate the displacement field\label{algo:step5}
  \begin{enumerate}
  \item Find the $M$ nearest\footnote{Alternatively, use samples inside a local geodesic ball, e.g. within the 1-ring neighbourhood  \cite{botsch2010polygon}.} samples to $\mathbf{x}_q$\label{algo:step5a}
  \item Compute $L_{m,k}\equiv L(\mathbf{d}_{m,k}\!\mid\! y(\mathbf{c}_m),\mathbf{s}_m(\mathbf{x}_m,\boldsymbol{\delta}_m)) = \sum_g L(y(\mathbf{c}_m)\!\mid\! g)R(\mathbf{x}_m+\mathbf{d}_{m,k}, g, \boldsymbol{\delta}_m)$\\to obtain a table where $m\in\{1,\ldots,M\}$ and $k\in\{1,\ldots, N_\text{displacement}\}$.\label{algo:step5b}
  \item Normalize each row s.t. $\max_k L(\mathbf{d}_{m,k}\!\mid\! y(\mathbf{c}_m),\mathbf{s}_m(\mathbf{x}_m,\boldsymbol{\delta}_m)) = 1$ for each $m$\label{algo:step5c}
  \item Compute weights incorporating proximity and directional preference:\\$w_{m,k}(\mathbf{x}_q) \propto \text{proximity}(\mathbf{x}_q,\mathbf{x}_m)\times\left|\left<\mathbf{n}_{\mathbf{x}_q},\mathbf{d}_{m,k}\right>\right|$ and $\sum w_{m,k} = 1$\label{algo:step5d}
  \item Compute $\overline{L}(\mathbf{d}_{k}\!\mid\! y(\mathbf{c}_m),\mathbf{s}_m) =\sum_m w_{m,k} L_{m,k}$\label{algo:step5e}
  \item Let $\mathbf{d}_\text{MAP}(\mathbf{x}_q)=\mathbf{d}_{m,k^*}$ where $k^*=\argmax_k \overline{L}(\mathbf{d}_{k}\!\mid\! y(\mathbf{c}_m),\mathbf{s}_m)$\label{algo:step5f}
  \end{enumerate}
\item Apply smoothing to MAP displacement estimate. For example,\label{algo:step6}
  \begin{enumerate}
  \item Compute inverse distance weights $w_{p,q}$ for neighbour points $\mathbf{x}_p\in\mathcal{N}_{\mathbf{x}_q}$ s.t. $\sum w_{p,q}=1$
  \item Compute $\mathbf{d}'_\text{MAP}(\mathbf{x}_q)=\sum_{p} w_{p,q}\cdot \mathbf{d}_\text{MAP}(\mathbf{x}_p)$
  \end{enumerate}
\item Apply correction to surface vertex to minimize discrepancy\label{algo:step7}
  \begin{enumerate}
  \item Update $\mathbf{x}'_q\leftarrow \mathbf{x}_q-\mathbf{d}'_\text{MAP}(\mathbf{x}_q)$\label{algo:step7a}
  \end{enumerate}
\item Post-processing step: resolve conflicts, e.g. any surface patch intersection that may arise.
\end{enumerate}

\subsection{Illustration}\label{sect:illustration}
\begin{figure*}[!htb]
\centering
\includegraphics[width=125mm]{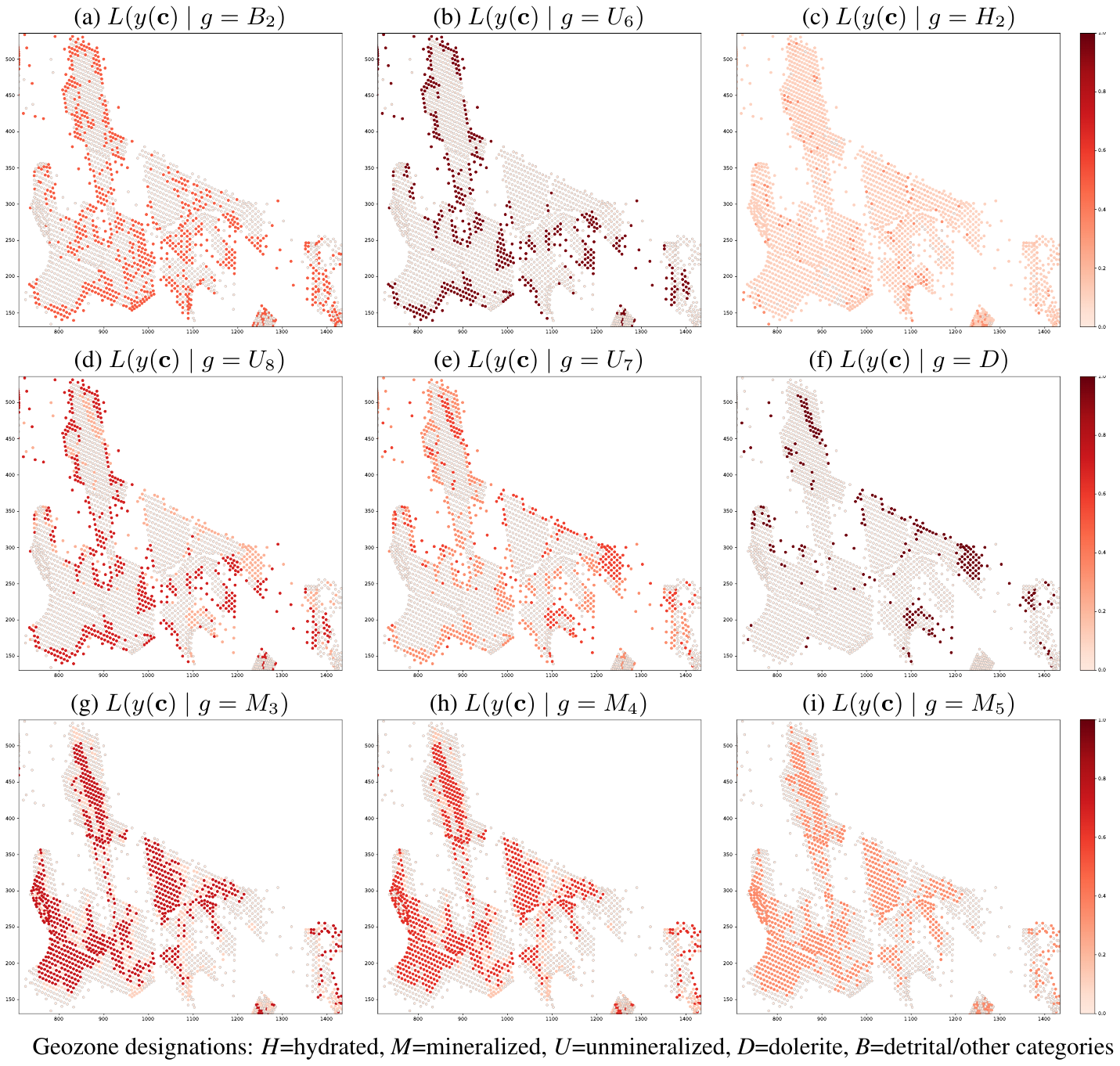}
\caption{Evaluation of $L(y(\mathbf{c}_i)\mid g)$ across geozones for many samples $\mathbf{c}_i$ within a spatial region of interest}
\label{fig:chemistry-likelihood-latex}
\end{figure*}

The Algorithm presented in Sec.~\ref{sect:spatial-warping-algorithm} step \ref{algo:step1} produces a table for $L(y(\mathbf{c})\mid g)$. The example shown in Fig.~\ref{fig:dest-tags-observation-likelihood-given-geozone} has size $N_\text{geozone}\!=\!24$, $N_\text{class}\!=\!7$. The exact definitions and geochemical mapping, $y(\mathbf{c})$, used will vary depending on the site. The main point to convey is the probabilistic association between chemistry and geozone, thus an assay sample with destination tag $y(\mathbf{c})$ has a plausible connection with multiple geozones in general. For instance, the HG tag has strong affinity with mineralized geozones (\textit{M}) but its association with hydrated domains (\textit{H}) cannot be discounted based on chemical evidence alone. This multivalent proposition is emphasized in Fig.~\ref{fig:chemistry-likelihood-latex} where each dot corresponds to a single blasthole assay observation $y(\mathbf{c}_i)$ (computed in step \ref{algo:step2}) and often a given location lit up in multiple geozones across different panels. This represents a spatial description of the evaluation in step \ref{algo:step3}. The panels in  Fig.~\ref{fig:chemistry-likelihood-latex} reveal spatial correlation with certain orebody structures and varying response (probability) associated with the dolerite dyke (\textit{D}) in row 2 and mineralized geozones (\textit{M}) in row 3.

\begin{table}[!htb]
\setlength\tabcolsep{0pt}
\begin{center}
\scriptsize
\begin{tabular}{c}
\includegraphics[width=95mm,trim={0mm 0mm 0mm 0mm},clip]{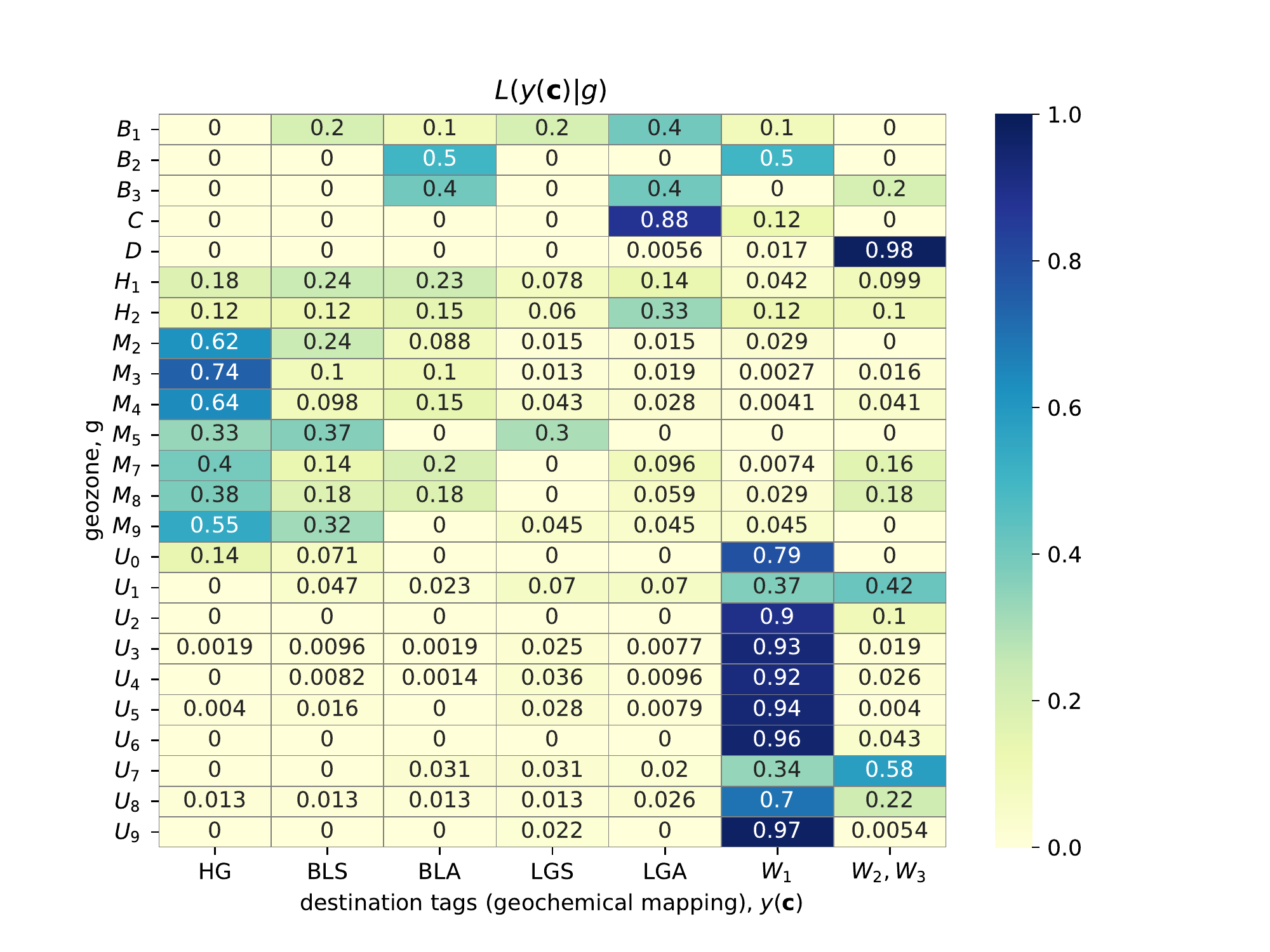}\\
Destination definitions --- HG (high grade): $\text{Fe}\ge 60$ and $\text{Al}_2\text{O}_3 < 6$, BL (blended): $\text{Fe}\in [55,60)$, LG (low grade): $\text{Fe}< 55$;\\
+S (siliceous: $\text{Al}_2\text{O}_3 < 3$), +A (aluminous: $\text{Al}_2\text{O}_3 \in[3,6)$), W\textsubscript{1} (waste: $\text{Fe}< 50$ and $\text{Al}_2\text{O}_3 < 6$), W\textsubscript{2,3} (waste: $\text{Al}_2\text{O}_3 \ge 6$)\\
Geozone definitions: \textit{H}=hydrated, \textit{M}=mineralized, \textit{U}=unmineralized, \textit{C}=canga, \textit{B}=detrital, \textit{D}=dolerite.
\end{tabular}
\end{center}
\captionof{figure}{Likelihood table of destination tag (geochemical mapping) given geozone, $L(y(\mathbf{c})\mid g)$}
\label{fig:dest-tags-observation-likelihood-given-geozone}
\end{table}

\begin{figure*}[!htb]
\centering
\includegraphics[width=140mm]{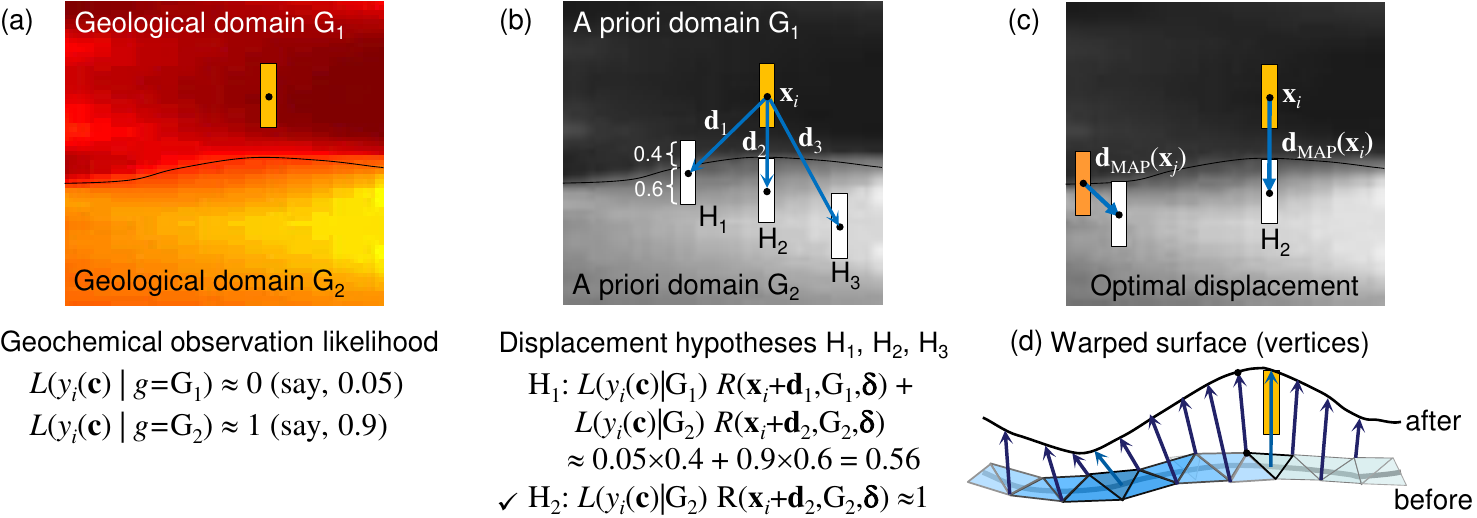}
\caption{Displacement estimation using $L(y(\textbf{c}_i)\mid g) R(\mathbf{x}_i+\mathbf{d}_i,g,\boldsymbol{\delta}_i)$}
\label{fig:geochemistry-spatial-likelihood}
\end{figure*}

Step~\ref{algo:step4} of the algorithm starts incorporating spatial information to disambiguate between geozone candidates whose chemistry are consistent with the observed assay sample. In concert, step~\ref{algo:step5} performs displacement estimation to minimize geochemical discrepancies with respect to the existing boundary represented by the surface. Fig.~\ref{fig:geochemistry-spatial-likelihood} provides a motivating example for both endeavors. First of all, if an assay sample is chemically consistent with the geozone it is currently situated in, there is no need for any spatial correction. It is only required when the observed chemistry is incongruent with the geochemical characteristics of the assumed domain at the measured location.

Figure~\ref{fig:geochemistry-spatial-likelihood}(a) shows an out-of-place (low grade Fe) observation within a mineralized domain. This is reflected numerically by $L(y(\mathbf{c}_i\mid G_1) \ll L(y(\mathbf{c}_i\mid G_2)$, where $G_1\in M$ and $G_2\in U$ belong to mineralized and unmineralized domains, respectively. Fig.~\ref{fig:geochemistry-spatial-likelihood}(b) elaborates on step~\ref{algo:step5b}: $L_{m,k}\equiv L(\mathbf{d}_{m,k}\!\mid\! y(\mathbf{c}_m),\mathbf{s}_m(\mathbf{x}_m,\boldsymbol{\delta}_m))$ for just one sample ($m=1$) and considers three displacement hypotheses $H_k$ with displacement vectors $\mathbf{d}_k\equiv\mathbf{d}_{m,k}, k\in\{1,...,K\}$ where $K=3$. The geozone-displacement likelihood $R(\mathbf{x}_m+\mathbf{d}_{m,k}, g, \boldsymbol{\delta}_m)$ is determined by the fraction of sample interval overlap with the a priori geozone structure. The example in Fig.~\ref{fig:geochemistry-spatial-likelihood}(b) shows the displacement hypothesis $H_2$ has the greatest support numerically. Intuitively, it represents the translation required to move the low grade sample completely into the unmineralized domain, $G_2$. Although hypothesis $H_3$ also provides a feasible solution, the displacement is greater and therefore it is not preferred.

\begin{figure*}[!htb]
\centering
\includegraphics[width=65mm]{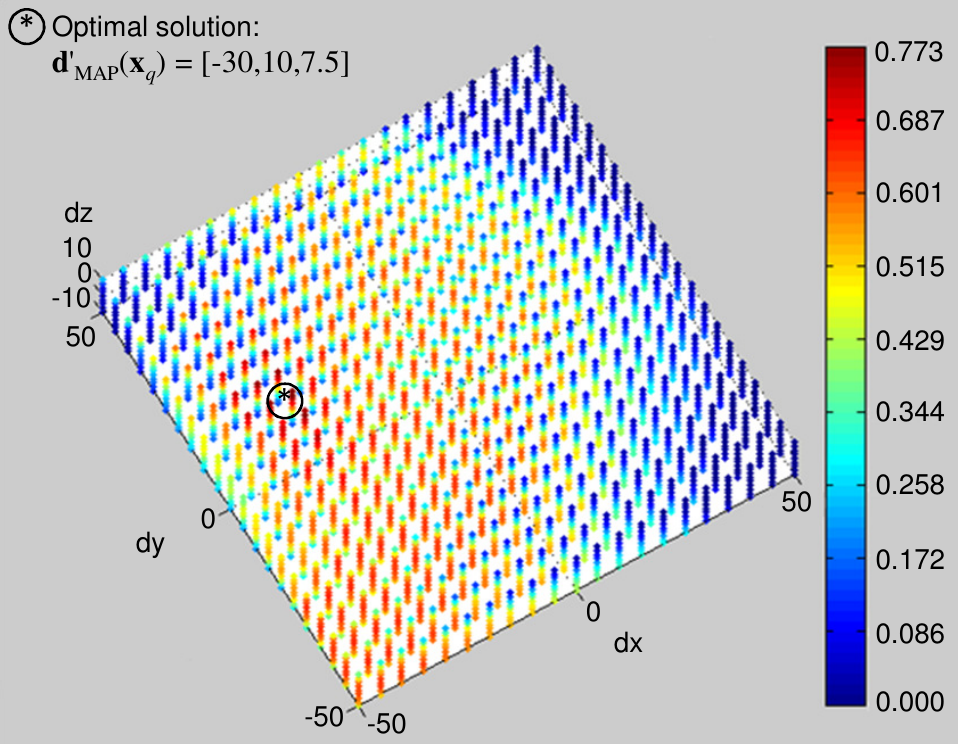}
\caption{The likelihood associated with each displacement vector in the parameter space. The optimal solution (see asterisk) represents the consensus amongst sample estimates $\mathbf{d}_\text{MAP}(\mathbf{x}_p)$ in the neighborhood of $\mathbf{x}_q$, $\mathcal{N}_{\mathbf{x}_q}$}
\label{fig:displacement-error-extrapolated}
\end{figure*}

Based on these principles, Fig.~\ref{fig:geochemistry-spatial-likelihood}(c) shows the optimal displacement estimated for $M=2$ samples. Put simply, step~\ref{algo:step5e} roughly corresponds to Fig.~\ref{fig:geochemistry-spatial-likelihood}(d) where the displacement computed for a surface vertex $\mathbf{x}_q$ is essentially a weighted average of the estimated  displacement from its $M$ nearest neighbors. Step~\ref{algo:step6} typically produces a locally smooth displacement error field. For an arbitrary vertex $\mathbf{x}_q$, the optimal solution, $\mathbf{d}'_\text{MAP}(\mathbf{x}_q)$, appears as a peak (see asterisk) in the parameter space in Fig.~\ref{fig:displacement-error-extrapolated}. Finally, the surface vertices are adjusted to complete step~\ref{algo:step7}, this may seen as additional smoothing or extrapolation on top of Fig.~\ref{fig:geochemistry-spatial-likelihood}(d).\footnote{In step~\ref{algo:step7a}, the update equation $\mathbf{x}'_q\leftarrow \mathbf{x}_q-\mathbf{d}'_\text{MAP}(\mathbf{x}_q)$ contains a minus sign because the displacement error estimation process considers moving samples around the initial boundary. In practice, we need to do the opposite, viz. move the boundary with respect to the samples which remain fixed at their measured locations.}

\subsection{Complexity}\label{sect:complexity}
\begin{figure*}[!htb]
\centering
\includegraphics[width=140mm]{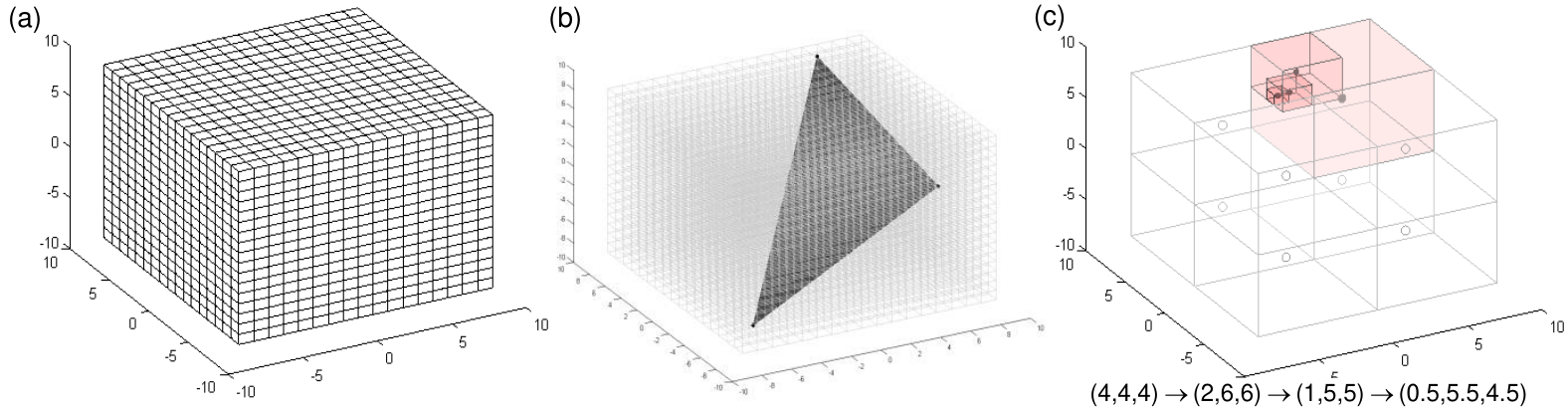}
\caption{Displacement search space exploration strategies: (a) brute-force, (b) subject to autoregressive or motion predictive modelling constraints, (c) hierarchical approach.}
\label{fig:complexity}
\end{figure*}

The main computational complexity lies in step~\ref{algo:step5} of the algorithm. Looking at $L_{m,k} = \sum_g L(y(\mathbf{c}_m)\!\mid\! g)R(\mathbf{x}_m+\mathbf{d}_{m,k}, g, \boldsymbol{\delta}_m)$, since $L(y(\mathbf{c}_m)\!\mid\! g)$ amounts to an $O(1)$ lookup operation, the per-sample complexity comes from $R(\mathbf{x}_m+\mathbf{d}_{m,k}, g, \boldsymbol{\delta}_m)$ where sample $m$ is fixed and $k\in\{1,\ldots,N_\text{displacement}\}$ varies over the displacement search space. Assuming a rectilinear, uniformly quantized search space over a lattice $\mathcal{L}\subset\mathbb{R}^3$ and $n$ grid points in each linear dimension, an exhaustive brute-force search (see Fig.~\ref{fig:complexity}a) is $O(n^3)$.

Depending on the surface geometry and sampling density, if the displacement estimation process is amenable to autoregressive or motion predictive modelling \cite{elnagar1998motion,kim2014robust}, complexity drops to $O(n^2)$. As an example, if the optimal displacement at three points in the vicinity of $\mathbf{x}_m$ are represented by the vertices of the triangle in the solution space in Fig.~\ref{fig:complexity}(b) and the cost function is locally convex, then the search may be constrained to the triangular region $\Omega\subset\mathbb{R}^2$. Although this paper does not mandate a specific implementation, it is worth noting that a hierarchical search strategy (see k-step successive refinement in Fig.~\ref{fig:complexity}c or \cite{konrad2005}) has an approximate complexity of $O(n\log n)$ since an ($2^3+1$) point search is conducted $k$ times, where $k\in\mathbb{Z}\approx \log_2 n$.

\section{Performance evaluation}\label{sect:boundary-warping-fusion-workflow-validation}
The benefits of spatial warping is first demonstrated, this will be followed by results from a large scale validation experiment.

\begin{figure*}[!htb]
\centering
\includegraphics[width=140mm]{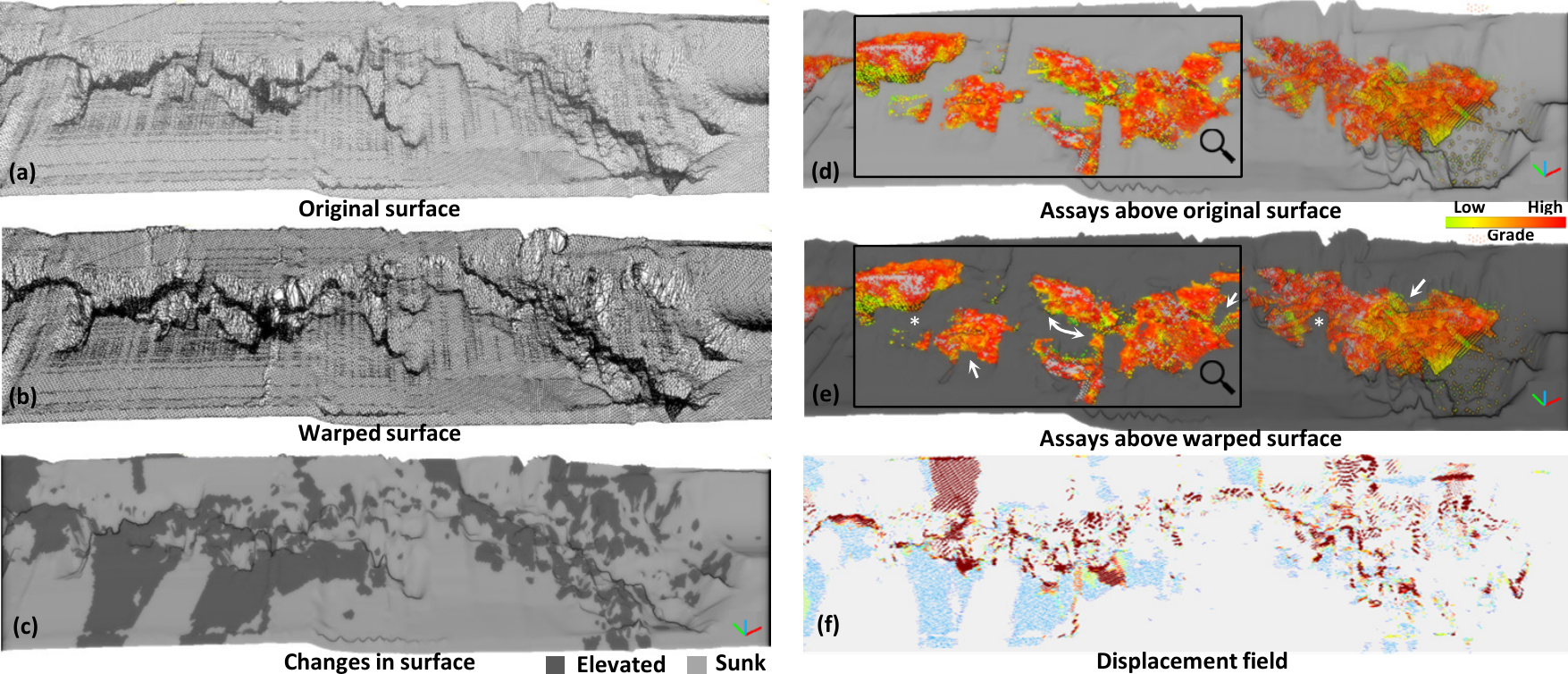}
\caption{Surface warping performed on a mineralization base surface. (a) original mesh surface, (b) warped surface, (c) map shows where the surface has elevated and sunk after warping. (d)--(e) Assays rendered above the original (resp. warped) surface are predominantly high-grade; (f) displacement field obtained through surface warping (an equivalent high resolution display is shown in Fig.~\ref{fig:displacement-field})}
\label{fig:surface-warping}
\end{figure*}

\subsection{Local corrections due to spatial warping}\label{sect:local-correction-spatial-warping}
Figure~\ref{fig:surface-warping} provides an overview of the surface warping result for a mineralization base surface where high grade material ideally sits above the boundary. The bottom panels illustrate the displacement field and highlight changes in elevation by superimposing the surfaces before and after warping. The right hand side panels in the top and middle row show the mineral (Fe) grade of the assay samples situated above the original and warped surfaces.

\begin{figure*}[!htb]
\centering
\includegraphics[width=125mm]{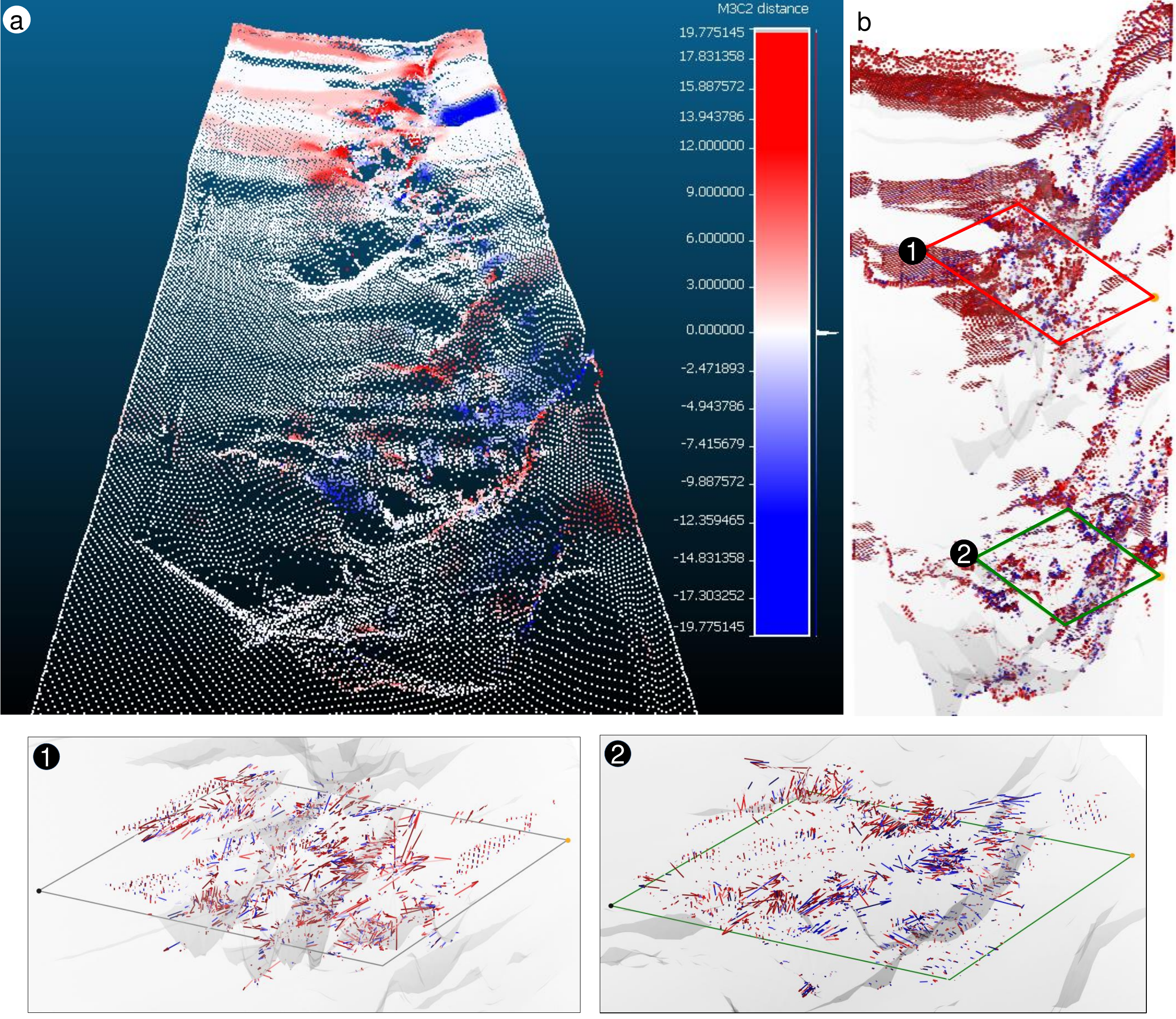}
\caption{Changes to the surface: (a) signed distance visualization obtained from (b) vertices displacement field; (c)--(d) magnified view of the displacement field in two sub-regions.}
\label{fig:displacement-field}
\end{figure*}

Figure~\ref{fig:displacement-field}(a) describes the topological changes to the warped surface. It conveys the same information as Fig.~\ref{fig:surface-warping}(c) albeit at higher resolution. This signed distance visualization is generated using a multiscale cloud compare algorithm \cite{lague2013accurate} which takes the original and warped surface vertices as input and displays elevation differences as a color map. Figure~\ref{fig:displacement-field}(b) and insets show the raw displacement field obtained directly from the warping process as a quiver plot.

\begin{figure*}[!htb]
\centering
\includegraphics[height=78mm,width=125mm]{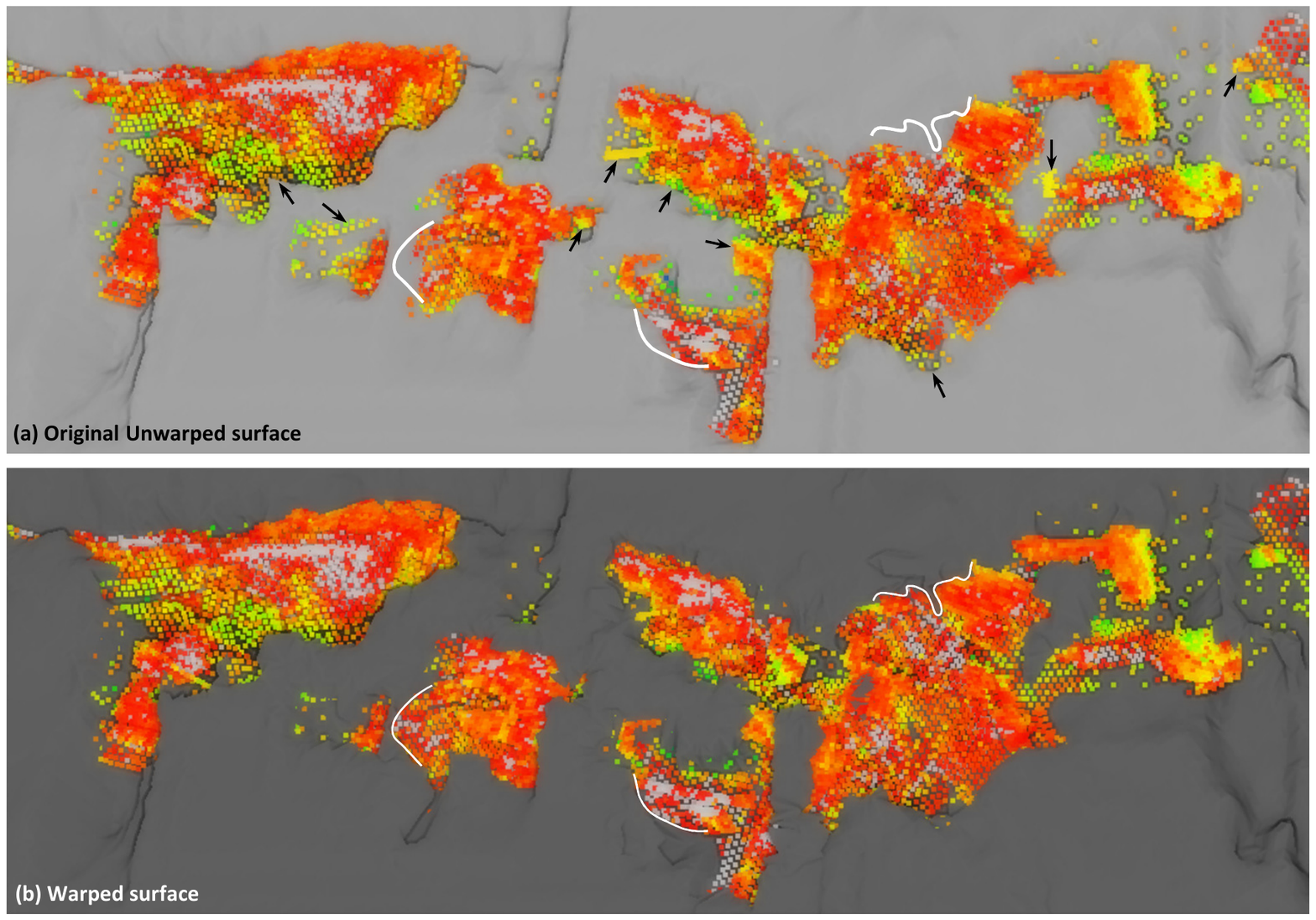}
\caption{Fe grade of assay samples situated above the (top) original and (bottom) warped `min\_base' surface.}
\label{fig:surface-warping-differences-magnified}
\end{figure*}

Figure~\ref{fig:surface-warping-differences-magnified} examines the effects of surface warping in more detail. It provides a magnified view of the rectangular region shown in Fig.~\ref{fig:surface-warping}(d) and (e). As expected, areas pointed by an arrow have contracted after surface warping. This results in the correct behaviour whereby non-mineralized samples have vanished below the mineralization boundary which is implicitly represented by the warped surface. The white lines indicate inclusive behaviour (or areas of expansion) where mineralized samples have risen above the mineralization boundary following surface warping. Overall, local delineation between high-grade and low-grade material has improved. The surface has effectively been pushed down to include more mineralized (red) samples and lifted up to exclude more low-grade (yellow) samples. Although the assay samples are coloured by iron grade alone in this illustration, it is worth bearing in mind that ore grade is assessed in practice as a function of multiple chemical components and this often includes Al\textsubscript{2}O\textsubscript{3} and other trace elements.

\begin{figure*}[!htb]
\centering
\includegraphics[width=125mm]{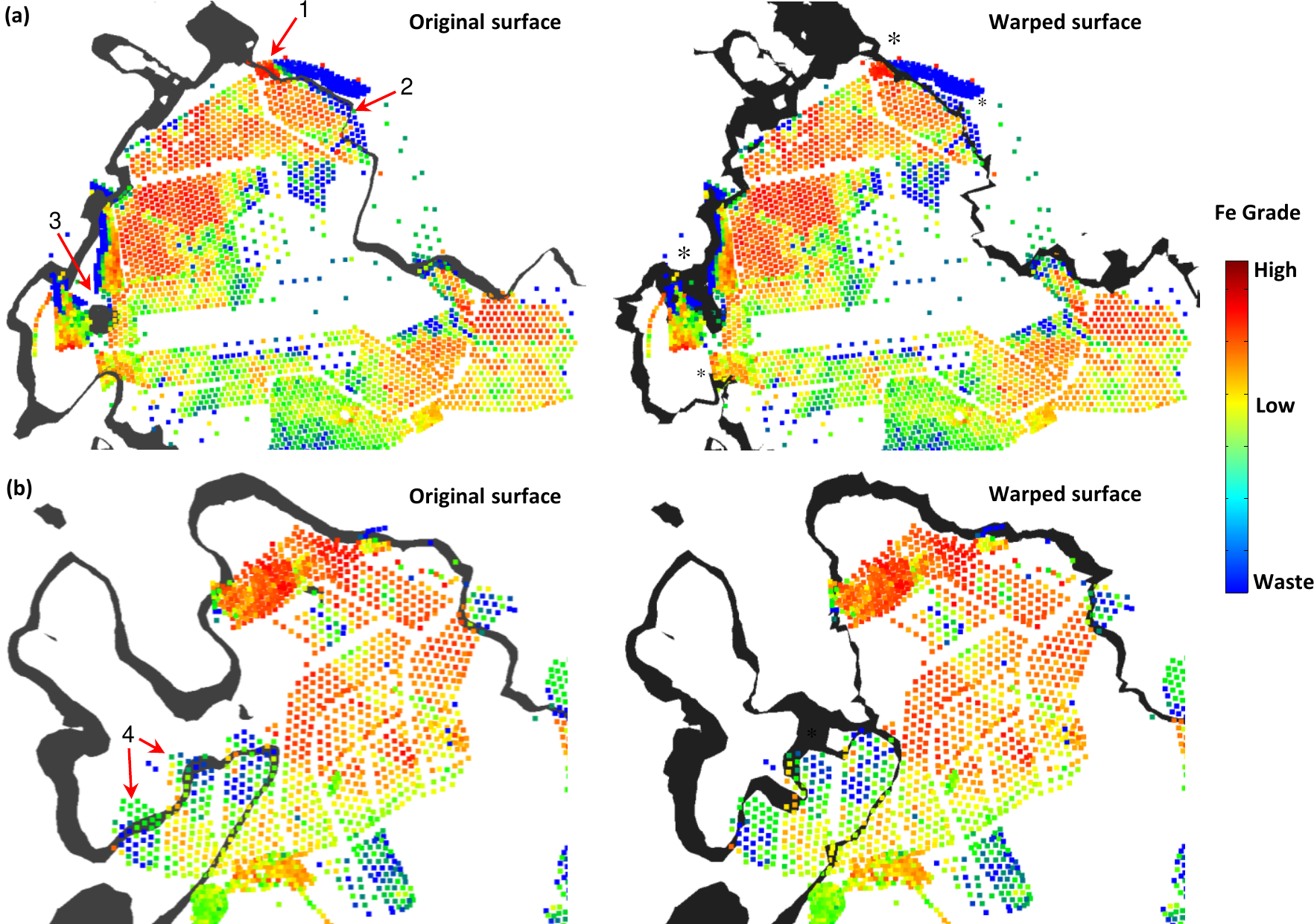}
\caption{Surface cross-sections taken at (a) $400\pm 5$m and (b) $440\pm 5$m show better delineation between waste and high-grade samples after warping.}
\label{fig:surface-sections-with-assays}
\end{figure*}

Figure~\ref{fig:surface-sections-with-assays} presents an alternative view. Surface cross-sections (the shell being visualized as black ribbons) are shown together with assay samples (coloured by grade) at two different elevations. Local changes are indicated by the arrows. The original surface is inaccurate at (a.1) as it excludes certain high-grade samples whereas in (a.2), (a.3) and (b.4) waste samples are included inside the boundary. Evidently, the warped surface provides better delineation between waste and high-grade samples as the boundary encircles the waste; this is especially noticeable at (b.4) in the warped surface. These observations can be verified quantitatively in Table~\ref{tab:samples-above-below-original-and-warped-surface} which demonstrates more effective separation of high grade and waste samples with warping.\footnote{Note that the dominant samples (HG/BL above the surface, and LG/W below the surface) do not quite reach 100\%. In part, this is due to some assay samples being taken from hole intervals that span across mineralized and non-mineralized geozones. Other surfaces that further delineate HG and W materials are not considered for the purpose of this evaluation.}

\begin{table}[h]
\begin{center}
\small
\setlength\tabcolsep{4pt}
\caption{Assay samples above and below the original and warped surfaces, categorised by High grade (HG), Blended (BL) siliceous and aluminous, Low grade (LG) and Waste (W).}\label{tab:samples-above-below-original-and-warped-surface}
\begin{tabular}{|l|p{12mm}|p{6mm}p{6mm}|p{6mm}p{6mm}|p{6mm}p{6mm}p{6mm}|c|c|}\hline
\multicolumn{11}{|c|}{Samples located above surface}\\ \hline
Surface & HG & BLS & BLA & LGS & LGA & W\textsubscript{1} & W\textsubscript{2} &  W\textsubscript{3} & $\frac{\text{(HG+BL)}}{\text{(HG+BL+LG+W)}}$ & $\frac{\text{(HG+BL)}}{\text{(HG+BL+W)}}$ \\ \hline
\textbf{original} & 12809 & 1609 & 3127 & 714 & 884 & 2588 & 1899 & 113 & 73.8\% & 79.2\% \\
\textbf{warped} & 13421 & 1730 & 3286 & 710 & 975 & 2194 & 1968 & 145 & 75.5\% & 81.0\% \\ \hline
change & ++++++ & + & ++ & & + & -\,-\,-\,- & + & & +1.7\% & +1.8\% \\ \hline
\multicolumn{11}{|c|}{Samples located below surface}\\ \hline
Surface & HG & BLS & BLA & LGS & LGA & W\textsubscript{1} & W\textsubscript{2} &  W\textsubscript{3} & $\frac{\text{(LG+W)}}{\text{(HG+BL+LG+W)}}$ & $\frac{\text{(W)}}{\text{(HG+W)}}$ \\ \hline
\textbf{original} & 1396 & 723 & 553 & 513 & 558 & 3874 & 876 & 223 & 69.3\% & 78.0\% \\
\textbf{warped} & 783 & 602 & 394 & 518 & 468 & 4268 & 806 & 191 & 77.8\% & 87.1\% \\ \hline
change & -\,-\,-\,-\,-\,- & - & -\,- & & - & ++++ & - & & +8.5\% & +9.1\% \\ \hline
\end{tabular}
\end{center}
\end{table}

\subsection{Validation experiment}\label{sect:validation-experiment}
To provide an objective evaluation, an end-to-end validation procedure known as r\textsubscript{2} spatial reconciliation is applied to assess the potential benefits of the proposed scheme where surface warping, block model restructuring and interval GP grade inferencing are applied, relative to a baseline resource model where none of these are used. The comparison requires computing r\textsubscript{2} values or ratios of (grade-block average)/(model predicted value) for each respective model and a chemical of interest, where a ``grade-block'' is an industry term that refers to regions with fairly constant composition and a typical volume ranging from 625\,m\textsuperscript{3} to 145,000\,m\textsuperscript{3}. These grade-blocks are marked with a destination tag for mining excavation purpose based on material types and/or the estimated grades. The grade-block averages (for Fe, SiO\textsubscript{2} etc.) are computed by geologists using the blast hole assays contained within the grade-block boundaries. The corresponding model predictions are volume-weighted averages of the GP inferenced mean grade values calculated over all blocks (perhaps numbered in the tens, hundreds or thousands) that intersect with each grade-block.\footnote{Fair sampling --- whether the number of samples taken is adequate and representative of the geology --- is an important consideration from the viewpoints of reliability and performance evaluation. In reality, suboptimal sampling does occur particularly in low-grade regions where the cost of extra sampling outweighs the benefit of knowing more about a waste zone with zero profit potential. In any event, the grade-block averages are as close to the ground truth as one can possibly attain.}

\begin{figure*}[!htb]
\centering
\includegraphics[width=140mm]{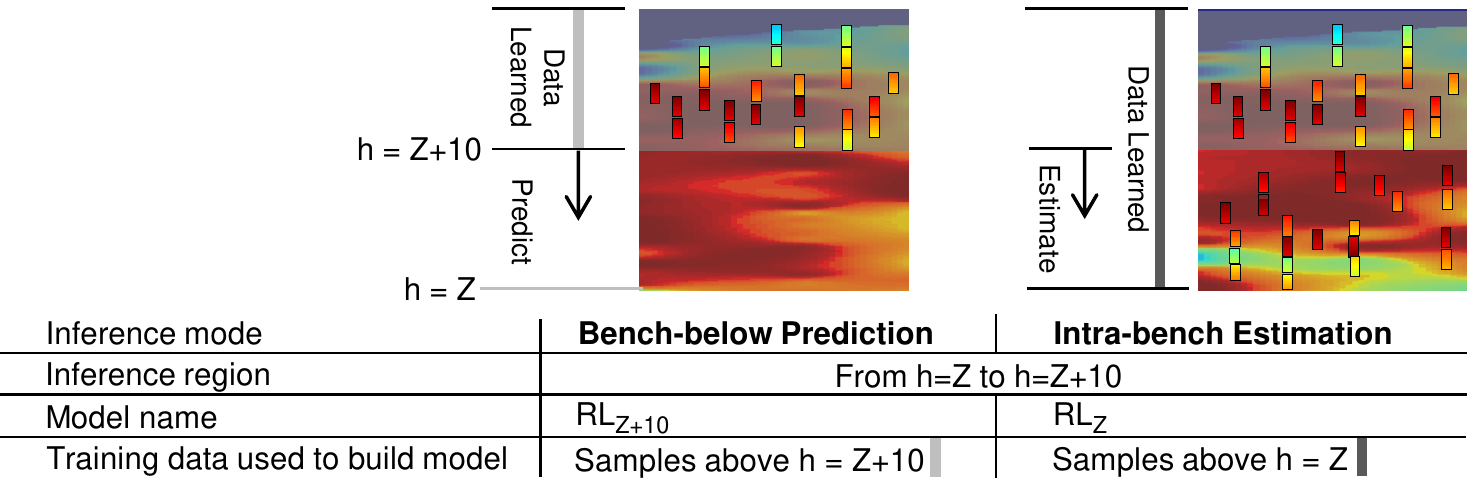}
\caption{Model evaluation inferencing modes. (Left) bench-below\,/\,forward prediction, (right) intra-bench estimation mode.}
\label{fig-intra-vs-inter-prediction}
\end{figure*}

Table~\ref{tab:r2-reconciliation-raw-data} shows the raw data for 5 grade-blocks and r\textsubscript{2} values computed for the proposed and reference model. Each grade-block is identified by the pit, bench and destination-tag. What is shown is only a snippet of a large table. The dest-tags\footnote{HG = high grade, WH = waste/hydrated, BGA = blended aluminous, LGS = low grade siliceous.} represent a classification based on average grade-block composition. For validation, we will consider two pits (A and B) from a Pilbara iron ore mine and five benches each of height 10m with a base elevation from 70m to 110m in 10m increment. In Table~\ref{tab:r2-reconciliation-raw-data}, bench 90 extends from a height of 90m to 100m. Accordingly, the grade-blocks used for evaluation are restricted to this z-interval. Model performance will be evaluated in `intra-bench' and 'predictive' mode. The former, indicated by RL\textsubscript{90} for instance, allows data down to a minimum elevation of 90m to be used during modelling (incl. GP training). Evaluation of an RL\textsubscript{90} model on bench 90 indicates how well a model interpolates the assay data. The latter, indicated by RL\textsubscript{100}, permits only data down to 100m to be used during modelling. Evaluation of an RL\textsubscript{100} model on unseen data from bench 90 focuses on a model's look-ahead (generalization and prediction) capability, viz. how well it vertically extrapolates the assay data. These differences are illustrated in Fig.~\ref{fig-intra-vs-inter-prediction}.

\begin{table}[h]
\begin{center}
\small
\setlength\tabcolsep{4pt}
\caption{r\textsubscript{2} spatial reconciliation: excerpt of raw data showing the grade-block (gb) and model predicted averages and normalized tonnage associated with 5 grade blocks.}\label{tab:r2-reconciliation-raw-data}
\begin{tabular}{|l|r|p{7mm}p{9mm}|p{8mm}|p{7mm}p{9mm}|p{8mm}|p{7mm}p{9mm}|p{8mm}|}\hline
\begin{tabular}{@{}l@{}}Pit\,/\,bench\,/\\ blast\#\,/\,dest-tag\end{tabular} & tonne\% & \begin{tabular}{@{}c@{}}Fe\\ (gb)\end{tabular} & \begin{tabular}{@{}c@{}}Fe\\ (model)\end{tabular} & \begin{tabular}{@{}c@{}}\textbf{Fe}\\ \textbf{r\textsubscript{2}}\end{tabular} & \begin{tabular}{@{}c@{}}SiO\textsubscript{2}\\ (gb)\end{tabular} & \begin{tabular}{@{}c@{}}SiO\textsubscript{2}\\ (model)\end{tabular} & \begin{tabular}{@{}c@{}}\textbf{SiO\textsubscript{2}}\\ \textbf{r\textsubscript{2}}\end{tabular} & \begin{tabular}{@{}c@{}}Al\textsubscript{2}O\textsubscript{3}\\ (gb)\end{tabular} & \begin{tabular}{@{}c@{}}Al\textsubscript{2}O\textsubscript{3}\\ (model)\end{tabular} & \begin{tabular}{@{}c@{}}\textbf{Al\textsubscript{2}O\textsubscript{3}}\\ \textbf{r\textsubscript{2}}\end{tabular}\\ \hline
\multicolumn{11}{|c|}{Proposed model}\\ \hline
A\,/\,90\,/\,1\,/\,HG13 & 1.47028 & 63.558 & 63.679 & 0.998 & 2.281 &\ 2.219 & 1.027 & 1.977 & 1.892 & 1.045\\
A\,/\,90\,/\,1\,/\,WH10 & 0.90878 & 47.969 & 54.935 & 0.873 & 25.786 & 14.664 & 1.758 & 1.762 & 2.022 & 0.871\\
A\,/\,90\,/\,3\,/\,HG23 & 1.25085 & 62.237 & 59.237 & 1.050 & 3.834 &\  7.195 & 0.532 & 2.011 & 1.758 & 1.143\\
A\,/\,90\,/\,399\,/\,BGA9 & 0.73250 & 56.618 & 55.832 & 1.014 & 8.807 &\ 9.748 & 0.903 & 3.034 & 3.228 & 0.940\\
A\,/\,90\,/\,5\,/\,LGS40 & 0.95260 & 54.531 & 52.953 & 1.029 & 9.440 & 10.383 & 0.909 & 6.853 & 7.581 & 0.904\\
\hline
\multicolumn{11}{|c|}{Reference model}\\ \hline
A\,/\,90\,/\,1\,/\,HG13 & 1.53145 & 63.558 & 62.942 & 1.009 & 2.281 &\ 2.640 & 0.863 & 1.977 & 2.098 & 0.942\\
A\,/\,90\,/\,1\,/\,WH10 & 0.89634 & 47.969 & 56.863 & 0.843 & 25.786 & 11.948 & 2.158 & 1.762 & 2.048 & 0.860\\
A\,/\,90\,/\,3\,/\,HG23 & 1.13243 & 62.237 & 63.299 & 0.983 & 3.834 &\ 2.434 & 1.575 & 2.011 & 2.061 & 0.975 \\
A\,/\,90\,/\,399\,/\,BGA9 & 0.63901 & 56.618 & 51.399 & 1.101 & 8.807 & 16.487 & 0.534 & 3.034 & 3.301 & 0.919 \\
A\,/\,90\,/\,5\,/\,LGS40 & 0.94149 & 54.531 & 56.785 & 0.960 & 9.440 &\ 8.183 & 1.153 & 6.853 & 5.719 & 1.198 \\
\hline
\end{tabular}
\end{center}
\end{table}

\begin{figure*}[!t]
\centering
\includegraphics[width=140mm]{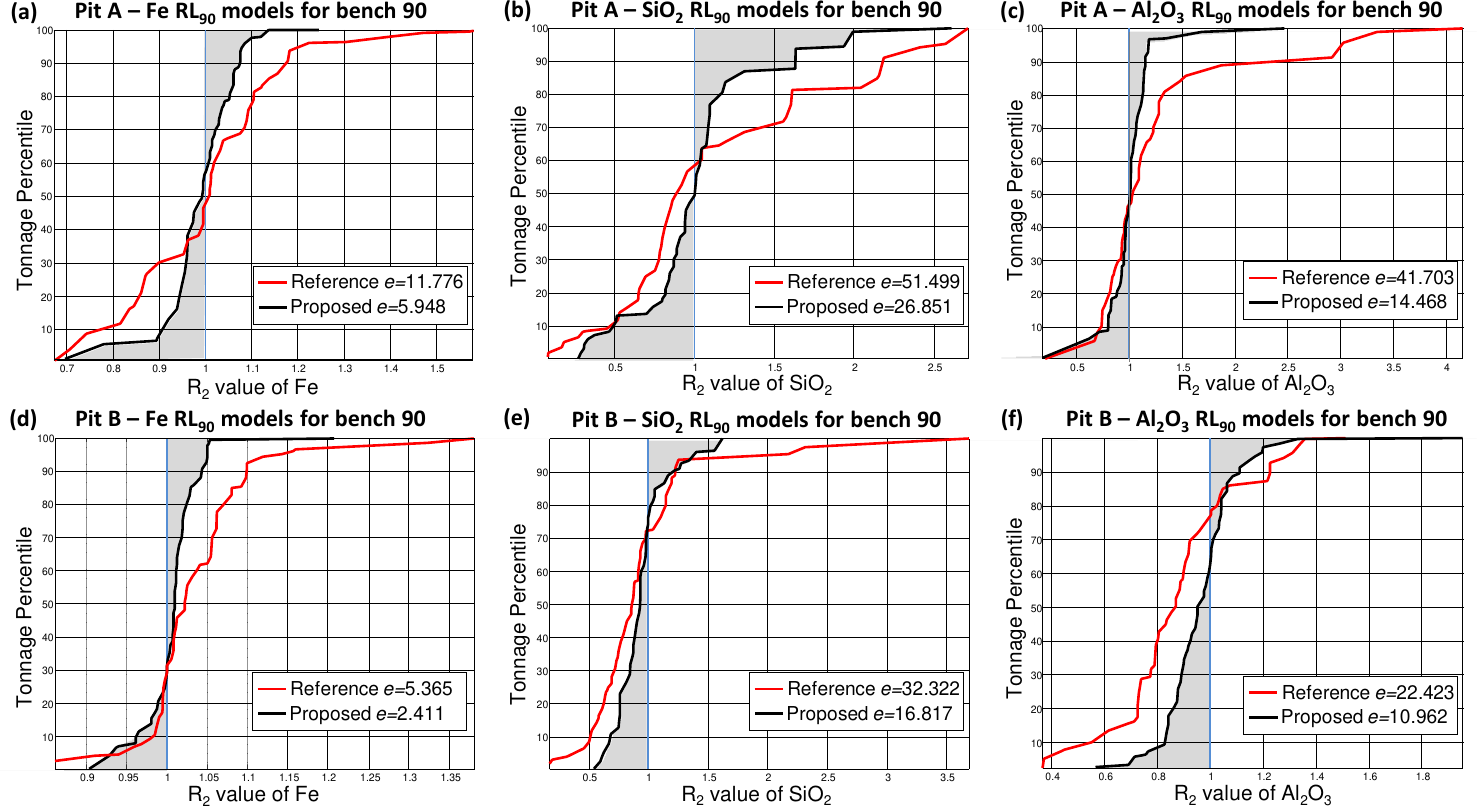}
\caption{Evaluation of orebody grade estimation RL\textsubscript{90} model using grade blocks from two pits and bench 90. The r\textsubscript{2} cumultative distribution function associated with pit A (top) and pit B (bottom) are shown from left to right for Fe, SiO\textsubscript{2} and Al\textsubscript{2}O\textsubscript{3}.}
\label{fig-obk-reconciliation-r2-cdf}
\end{figure*}

In order to convey useful information for large-scale performance evaluation, we propose using an r\textsubscript{2} error score. First, the r\textsubscript{2} values associated with bench $z$ and an  RL\textsubscript{\textit{h}} model (where $h=z$ in intra-bench mode, or $h=z+10$ in predictive mode) are sorted in increasing order and cumulative tonnage percentages are computed. This produces an r\textsubscript{2} cumulative distribution function (cdf); an example of which is shown in Fig.~\ref{fig-obk-reconciliation-r2-cdf}(a). An $r_2$ value less than 1 indicates over-estimation by the model, conversely, a value greater than 1 indicates under-estimation w.r.t. the grade-blocks. In a perfect scenario where there is zero discrepancy between the grade-blocks and model predicted values, the cdf curve becomes a step function that transitions from 0\% to 100\% at an r\textsubscript{2} value of 1. Hence, adding the area below the curve for $r_2 < 1$ to the area above the curve for $r_2 \ge 1$ provides an aggregate error measure (a performance statistic) of a model for a given pit, bench and chemical. The graphs also give insight. For instance, the left-leaning red curve in Fig.~\ref{fig-obk-reconciliation-r2-cdf}(f) provides evidence of bias, viz. the reference model has a tendency of over-estimating the Al\textsubscript{2}O\textsubscript{3} grade in pit B for bench 90.

\begin{table}[h]
\begin{center}
\footnotesize
\setlength\tabcolsep{4pt}
\caption{r\textsubscript{2} spatial reconciliation $r_2$ statistics.}\label{tab:r2-reconciliation-stats}
\begin{tabular}{|c|c|ccc|ccc|ccc|ccc|}\hline
& & \multicolumn{6}{c|}{Pit A} & \multicolumn{6}{c|}{Pit B}\\
& &\multicolumn{3}{c}{Proposed model} & \multicolumn{3}{c|}{Reference}& \multicolumn{3}{c}{Proposed model} & \multicolumn{3}{c|}{Reference}\\ \hline
Bench & Model & Fe & SiO\textsubscript{2} & Al\textsubscript{2}O\textsubscript{3} & Fe & SiO\textsubscript{2} & Al\textsubscript{2}O\textsubscript{3} & Fe & SiO\textsubscript{2} & Al\textsubscript{2}O\textsubscript{3} & Fe & SiO\textsubscript{2} & Al\textsubscript{2}O\textsubscript{3}\\ \hline
\multicolumn{2}{c}{} &\multicolumn{12}{c}{Intra-bench estimation performance}\\ \hline
110 & RL\textsubscript{110} & \textbf{4.271}  & \textbf{18.202}  & \textbf{12.059} & 7.761 & 34.531 & 19.995 & \textbf{3.351} & \textbf{20.905} & \textbf{17.516} & 4.069 & 21.737 & 20.087\\
100 & RL\textsubscript{100} & \textbf{5.411} & \textbf{28.101} & \textbf{24.784} & 8.548 & 46.368 & 29.377 & \textbf{2.093} & \textbf{15.670} & \textbf{11.505} & 3.282 & 22.298 & 19.057\\
90 & RL\textsubscript{90} & \textbf{5.948} & \textbf{26.851} & \textbf{14.468} & 11.776 & 51.499 & 41.703 & \textbf{2.411} & \textbf{16.817} & \textbf{10.962} & 5.365 & 32.322 & 22.423\\
80 & RL\textsubscript{80} & \textbf{4.693} & \textbf{16.172} & \textbf{21.967} & 12.182 & 60.436 & 27.069 & \textbf{6.777} & \textbf{22.986} & \textbf{18.957} & 10.461 & 39.828 & 29.820\\
70 & RL\textsubscript{70} & \textbf{9.420} & \textbf{19.646} & \textbf{25.036} & 11.135 & 49.092 & 53.319 & \textbf{9.416} & \textbf{24.409} & \textbf{44.289} & 9.688 & 39.862 & 77.711 \\ \hline
$\mu_\text{g}$ & & \textbf{5.711} & \textbf{21.280} & \textbf{18.828} & 10.117 & 47.611 & 31.816 & \textbf{4.042} & \textbf{19.862} & \textbf{17.933} & 5.919 & 30.140 & 28.823\\ \hline 

\multicolumn{2}{c}{} &\multicolumn{12}{c}{Bench-below prediction performance}\\ \hline
100 & RL\textsubscript{110} & \textbf{6.707} & \textbf{41.035} & 34.372 & 8.548 & 46.372 & 29.381 & 4.136 & 20.800 & 25.621 & 3.282 & 19.060 & 22.301 \\
90 & RL\textsubscript{100} & \textbf{9.832} & \textbf{41.214} & \textbf{26.107} & 11.776 & 51.495 & 38.455 & \textbf{4.100} & \textbf{29.440} & \textbf{17.766} & 5.365 & 32.322 & 22.425 \\
80 & RL\textsubscript{90} & \textbf{7.336} & \textbf{24.695} & \textbf{21.196} & 12.182 & 60.437 & 27.067 & \textbf{8.794} & \textbf{29.516} & \textbf{27.742} & 10.461 & 39.821 & 29.817 \\
70 & RL\textsubscript{80} & \textbf{8.069} & \textbf{28.985} & \textbf{42.950} & 11.135 & 49.097 & 53.319 & \textbf{9.605} & \textbf{25.907} & \textbf{63.121} & 9.688 & 39.859 & 77.786 \\ \hline
$\mu_\text{g}$ & & \textbf{7.904} & \textbf{33.170} & \textbf{30.064} & 10.810 & 51.594 & 35.734 & \textbf{6.152} & \textbf{27.558} & \textbf{28.362} & 6.500 & 32.705 & 31.554\\ \hline 
\multicolumn{14}{c}{$\mu_\text{g}$ denotes the geometric mean}
\end{tabular}
\end{center}
\end{table}

The two pits combined contain over 400 grade-blocks and a volume in excess of $3\times10^6$ m\textsuperscript{3}. The summary statistics for intra-bench estimation and bench-below prediction are shown in Table~\ref{tab:r2-reconciliation-stats}. The main observation is that the proposed model outperforms the reference model. With few exceptions, the r\textsubscript{2} error scores are consistently lower for both intra-bench and bench-below prediction for Fe, SiO\textsubscript{2} and Al\textsubscript{2}O\textsubscript{3}. Generally, the error scores are higher for bench-below prediction, the performance gap reflects the relative difficulty of the problem. These findings suggest the use of surface warping, block model spatial restructuring and interval GP inference can increase accuracy in grade estimation. In particular, improving the alignment of mesh surfaces with respect to the underlying boundaries (using the observed geochemistry from samples to enforce consistency) can make a real difference and improve orebody modelling outcomes.

\subsection{Discussion}\label{sect:discussion}
Although the results have been analyzed in the context of improving grade models, there is also intrinsic value in updating geological boundaries via surface warping using geochemical assay data. The ability to better characterize the lithological contacts (e.g. transition between mineralized and unmineralized domains) improves the mine geologist's ability to understand ore genesis events, particularly fluid flow within major structures in a banded iron formation (BIF) hosted iron ore sytem \cite{hagemann2016bif,perring2020new}. Shale boundaries\footnote{Shale is essentially a clay mineral with elevated SiO\textsubscript{2} and Al\textsubscript{2}O\textsubscript{3} content.}, along with folds and faults, play a significant role in controlling mineralization as fluid pathways are restricted by impermeable layers such as shale bands.

From a modelling perspective, warped surfaces have been used successfully to modify the block model structure in  \cite{leung2020mos} to improve the delineation between domains for grade estimation. Commercially available geo-modelling softwares often employ implicit modelling techniques \cite{cowan2003practical} to extract boundary as an iso-surface from a manifold. Recent work by Renaudeau et al.\,even handles discontinuities \cite{renaudeau2019implicit}. A recognized problem with this process is that it generally assumes the entire dataset is available and models everything at once. In contrast, our proposal honors the geologist's interpretations and uses the mesh surfaces prepared by experts as starting points. The warping procedure can be applied iteratively as more production data (for instance, blasthole assays from a single bench) become available to rectify surfaces incrementally. This flexible approach works well with sparse data and periodic updates which underscore the progressive nature of open-pit mining.

More accurate knowledge about these geological boundaries can be exploited to improve decision making and process efficiency. It facilitates dynamic mine planning and presents options on whether to excavate, postpone or abandon operations in an area. Then, there is the fleet management and task scheduling aspect which coordinates the movement of mining equipment \cite{samavati2019improvements} such as diggers, excavators and haul trucks, directing them when and where to go to achieve flow targets \cite{seiler2020flow}. Knowing where an ore/waste transition might occur, drill-charge-and-blast operations may be optimized with respect to placement and depth. The directional information conveyed by the surfaces also provides guidance for adaptive sampling \cite{ahsan-15} which maximizes utility and minimizes cost. This allows specific areas to be targeted where the boundary transition is most uncertain.

\begin{figure*}[!t]
\centering
\includegraphics[width=85mm]{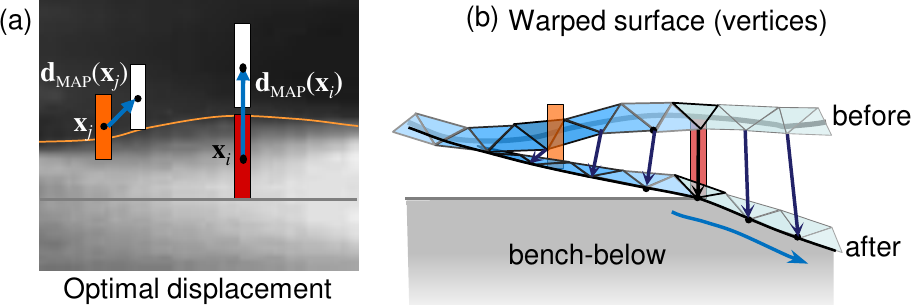}
\caption{Surface warping can propagate directional information about a boundary to the bench below}
\label{fig-changes-propagating-to-bench-below}
\end{figure*}

As Fig.~\ref{fig-changes-propagating-to-bench-below} shows, surface warping can project information to an adjacent bench or the bench below; this gives to some extent a forward stratigraphic modelling capability, allowing the drilling density and assay sampling rate to be adjusted if needed. Geological risks \cite{benndorf2013stochastic} can be mitigated with an improved grade model. These risks include identifying sticky or hazardous material, as well as grade control in general, such as finding waste instead of ore in a variable grade block.

\section{Conclusion}\label{sect:conclusion}
This paper described the importance of having an accurate surface for grade estimation in mining. As motivation, it was shown that an inaccurate surface --- one that fails to capture the location and shape of the underlying geological boundary --- can impact the block structure and inferencing ability of the resultant grade estimation model which in turn can lead to smearing and misleading interpretations. The main contribution was a Bayesian formulation of the surface warping problem which seeks to maximize the agreement between the surface and observed data. The objective was to reshape the surface where possible to provide a clear delineation between ore and waste material that is consistent with the observations. This involved estimating and applying the optimal displacement to the mesh vertices based on spatial and compositional analysis. The maximum a posteriori (MAP) solution considered the chemistry observation likelihood in a given geozone and incorporated an a priori spatial structure which allows the likelihood of a displacement estimate to be computed using geological domain knowledge. The results showed that locally, the mineralized and non-mineralized samples are better separated by the warped surface.

For end-to-end performance evaluation which encompasses surface warping, block model spatial restructuring, and grade estimation based on GP inferencing, the \textit{r\textsubscript{2} reconciliation error score} was proposed. This provided a grade model validation metric that is useful irrespective of the actual algorithms\,/\,processes deployed in the system components. Our experiments showed the r\textsubscript{2} error scores were consistently and significantly lower with surface warping when the estimated grades for chemicals of interest (Fe, SiO\textsubscript{2} and Al\textsubscript{2}O\textsubscript{3}) were compared with over 400 grade-blocks for two large pits, having a total volume in excess of $3\times 10^6$ m\textsuperscript{3}. This demonstrated the value of implementing the displacement estimation framework for surface warping\,/\,boundary rectification, and spatial algorithms more generally, in a mining automation context.

\section{Authorship statement}
The first two authors contributed to the core contents in this paper. Alexander Lowe studied surface warping (mesh vertices displacement estimation) as a maximum likelihood problem. He investigated different strategies and provided a concrete software implementation. Raymond Leung reformulated the problem in a Bayesian framework, developed the block model spatial restructuring algorithms, and proposed using the R\textsubscript{2} CDF error score as a validation measure. He conceptualized this paper and wrote the manuscript in consultation with other authors. Anna Chlingaryan and Arman Melkumyan devised the covariance functions and mathematical framework for grade estimation and GP inference using interval data (see Appendix). John Zigman conducted the validation experiments. John and Raymond performed data analysis and interpreted the results.

\begin{appendix}
\section{Appendix: Gaussian Processes --- inferencing for grade estimation}\label{sect:inferencing-for-grade-estimation}
Inferencing refers to the task of predicting the grade value for certain chemicals of interest at locations where direct assay measurements are unavailable. In this exposition, Gaussian Process (GP) is used to provide a probabilistic model of the grade functions given a set of data, this allows both the mean and uncertainty associated with the compositional percentage of various chemicals such as Fe, SiO\textsubscript{2} etc.\,to be estimated. Inferencing is generally preceded by a training phase which optimizes the hyper-parameters that describe the Gaussian Process.

Mathematically, a GP is an infinite collection of random variables, any finite number of which has a joint Gaussian distribution. Machine learning using GPs consists of two steps: training and inference. For training, simulated annealing and gradient descent procedures are used to optimize the hyper-parameters to create a probabilistic model that best represents the training data. Specifically, the GP hyper-parameters include \textit{length scales} that describe the rate of change in composition with space, and \textit{noise variance} that describes the amount of noise present in the data.

A training set $\mathcal{T}=(X,\mathbf{y})$ consists of a matrix of training samples $X=[x_1,x_2,\ldots,x_N]^T\in\mathbb{R}^{N\times D}$ and corresponding target vector $\mathbf{y}=[y_1,y_2,\ldots,y_N]\in\mathbb{R}^N$. Here, $N$ represents the number of training samples available for a geozone. Each $x_i\in\mathbb{R}^D$ denotes an observation (the spatial coordinates where an assay sample is taken) and the associated value $y_i\in\mathbb{R}$ denotes a chemical's compositional percentage. The objective is to compute the predictive distribution $f(x_*)$ at various test points $x_*$.  Formally, a GP model places a multivariate Gaussian distribution over the space of function variables $f(x)$, mapping input to output spaces. GPs can also be considered as a stochastic process that can be fully specified by its mean function $m(x)$ and covariance function $k(x,x')$. To completely describe the standard regression model, we assume Gaussian noise $\varepsilon$ with variance $\sigma_\text{n}^2$, so that $y=f(x)+\varepsilon$. With a training set $(X, f, y)=(\{x_i\}, \{f_i\}, \{y_i\})_{i=1:N}$ and test set $(X, f, y)=(\{x_{*i}\}, \{f_{*i}\}, \{y_{*i}\})_{i=1:N}$ where $\{y_i\}$ are observed and $\{y_{*i}\}$ are unknown and $m(x)=0$, the joint distribution becomes
\begin{align}
\begin{bmatrix}y \\ f_*\end{bmatrix}\sim \mathcal{N}\left(0,\begin{bmatrix}K(X,X)+\sigma_\text{n}^2 I & K(X,X_*)\\K(X_*,X)& K(X_*,X_*)\end{bmatrix}\right)
\label{eq:gp-joint-distribution}
\end{align}
In Equation~(\ref{eq:gp-joint-distribution}), $\mathcal{N}(\mu, cov(f_*))$ is a multivariate Gaussian distribution with mean $\mu$, posterior  covariance at the estimated locations $cov(f_*)$, and $K$ is the covariance matrix computed between all the points in the set. Thus, the matrix element $K_{i,j*}\equiv K(X_i,X_{*j})$ for instance is obtained by applying the kernel to the locations of sample $x_i$ and $x_{*j}$ from the training and test sets, respectively. By conditioning on the observed training points, the predictive distribution for new points can be obtained as:
\begin{align}
p(f_*\!\mid\! X_*,X,y)=\mathcal{N}(\mu, cov(f_*))\label{eq:gp-predictive-dist}
\end{align}
where
\begin{align}
\mu&=K(X_*,X)\left[K(X,X)+\sigma_\text{n}^2I\right]^{-1} y\label{eq:gp-predictive-mean}
\end{align}
and the posterior covariance
\begin{align}
cov(f_*)&=K(X_*,X_*) - K(X_*,X)\left[K(X,X)+\sigma_\text{n}^2 I\right]^{-1} K(X,X_*)\label{eq:gp-predictive-covariance}
\end{align}
Learning a GP model is equivalent to learning the hyper-parameters of the covariance function from a data set. In a Bayesian framework, this can be performed by maximising the log of the marginal likelihood \cite{williams2006gaussian} with respect to $\theta$:
\begin{align}
\log p(y\mid X,\theta)=-\frac{1}{2}y^T\left[K(X,X)+\sigma_\text{n}^2 I\right]^{-1}y -\frac{1}{2}\log\left|K(X,X)+\sigma_\text{n}^2 I\right|-\frac{N}{2}\log 2\pi\label{eq:gp-nlml}
\end{align}
The marginal likelihood is a non-convex function, thus only local maxima can be obtained. It has three terms (from left to right) that represent the data fit, complexity penalty (to include the Occam's razor principle) and a normalization constant. In this standard framework, the main contribution is the design of new kernels \cite{rtgi-rtcma2014} that deal with not only point-based observations, but also \textit{interval} observations where $y_i$ represents an average assay value measured over some interval in drilled holes. This enables performing data fusion between exploration and blast hole assays taking into consideration their respective supports. The GP covariance functions for interval data are described mathematically in \cite{rtgi-rtcma2014}. Relevant works that underpin this theory can be found in \cite{jewbali2011apcom} and \cite{rasmussen2003bayesian}.
\end{appendix}

\vspace{3mm}
\section*{Acknowledgment}
This work was supported by the Australian Centre for Field Robotics and the Rio Tinto Centre for Mine Automation.

\bibliographystyle{unsrt}  
\bibliography{ms}

\end{document}